\documentclass[a4paper,11pt]{article}
\usepackage{jheppub}
\usepackage{amsmath,amssymb}
\usepackage{graphicx}
\usepackage{tikz}
\usepackage{longtable}
\usepackage{tabularx}
\usepackage{float}
\usepackage{makecell}
\usepackage{booktabs}
\usepackage{tikz-feynman}
\usepackage{tabularx}
\usepackage{multirow}
\usepackage{array}

\newcommand{\onubb}{0\nu\beta\beta}
\newcommand{\SMEFT}{\mathrm{SMEFT}}
\newcommand{\LEFT}{\mathrm{LEFT}}
\newcommand{\LNV}{\Delta L=2}
\newcommand{\SU}[1]{\mathrm{SU}(#1)}

\newcommand{\gev}{\mathrm{GeV}}

\title{Complete UV Resonances of SMEFT Dim-9 Operators for Short-range Neutrinoless Double Beta Decay}

\author[e]{Hao-Lin Li, }
\author[g]{Yu-Han Ni, }
\author[d]{Ming-Lei Xiao, }
\author[a,b,c,f]{Jiang-Hao Yu, }
\author[a,b,c]{Xiao-Long Zheng, }

\affiliation[a]{School of Fundamental Physics and Mathematical Sciences, Hangzhou Institute for Advanced
Study, UCAS, Hangzhou 310024, China}
\affiliation[b]{Institute of Theoretical Physics, Chinese Academy of Sciences, Beijing 100190, China}
\affiliation[c]{School of Physical Sciences, University of Chinese Academy of Sciences, Beijing 100049, P.\ R.\ China}
\affiliation[d]{School of Science, Sun Yat-Sen University, Shenzhen 518107, P. R. China}
\affiliation[e]{School of Physics, Sun Yat-Sen University, Guangzhou 510275, P. R. China}
\affiliation[f]{International Center for Theoretical Physics Asia-Pacific, Beijing/Hangzhou, China}
\affiliation[g]{School of Science and Engineering, The Chinese University of Hong Kong, Shenzhen, Shenzhen 518172, China}

\emailAdd{lihlin68@mail.sysu.edu.cn}
\emailAdd{niyuhan@cuhk.edu.cn}
\emailAdd{xiaomlei@mail.sysu.edu.cn}
\emailAdd{jhyu@itp.ac.cn}
\emailAdd{zhengxiaolong23@mails.ucas.ac.cn}

\abstract{

We present a systematic classification of tree-level ultraviolet (UV) completions for dimension-nine SMEFT operators relevant to short-range neutrinoless double beta decay ($0\nu\beta\beta$). Using the SMEFT J-basis framework, we categorize distinct UV completions, including both all-boson and boson-fermion-boson topologies. A primary objective is the identification of \emph{minimal UV realizations}, defined as the smallest set of genuine heavy degrees of freedom required to generate each operator. Out of the 505 unique mediator combinations identified, 440 are found to be minimal, with 12 cases necessitating only two distinct heavy species.
While our findings reproduce the scalar- and fermion-mediated results of Ref.~\cite{Bonnet:2012kh}, we extend earlier studies of vector representations~\cite{Fonseca:2016tbn} by providing a complete compilation of 324 minimal UV completions featuring vector resonances within the short-range dim-9 SMEFT classification.

}

\begin{document}
\maketitle
\flushbottom
\allowdisplaybreaks

\section{Introduction}
\label{sec:intro}

Neutrinoless double beta decay ($\onubb$) is a highly sensitive probe of lepton-number violation (LNV) and the Majorana nature of neutrinos~\cite{Zeldovich:1981, Schechter:1981bd,Rodejohann:2011mu,Deppisch:2012nb,Dolinski:2019nrj,Engel:2016xgb}.
Observing this transition would establish LNV dynamics beyond the Standard Model (SM) and help explain  the origin of neutrino masses~\cite{Agostini:2022zub}.
From the perspective of effective field theory (EFT), the underlying physics is organized systematically in an expansion of operator dimensions~\cite{Weinberg:1979sa,Babu:2001ex,deGouvea:2007xp,Bonnet:2009ej,Lehman:2014jma,Liao:2016hru,Liao:2020jmk}, with the dimension-seven sector understood in its complete nonredundant form~\cite{Lehman:2014jma,Liao:2016hru}, and long-range mechanisms separated from genuinely short-range contributions encoded in local contact operators~\cite{Prezeau:2003xn,Pas:1999fc,Pas:2000vn,Cirigliano:2018hja}.
The latter provide a particularly direct bridge between experimental half-life limits and ultraviolet (UV) dynamics.
They also raise a practical question: how can one organize the short-distance operators, together with the heavy mediators that generate them, in a gauge-invariant way before electroweak symmetry breaking?

In many UV scenarios, the dominant short-range contribution is encoded in dimension-nine interactions involving two electrons and four quarks.
Their matching and renormalization group evolution down to hadronic scales can be treated systematically in the low-energy EFT~\cite{Cirigliano:2018yza,Cirigliano:2017djv,Dekens:2020ttz,Gonzalez:2015ady,Cirigliano:2022oqy,Cirigliano:2021qko}.
However, constructing a complete description of the dimension-nine sector in the unbroken phase is technically subtle.
Once electroweak invariance is imposed, six-fermion structures admit several inequivalent color and weak-isospin contractions, and accounting for these variations becomes complex in the presence of repeated fields, permutation symmetry, and SM-like contributions.
Moreover, the UV interpretation of a ``tree-level completion'' depends on explicit power counting.
A mediator exchange that is tree-level in a diagrammatic sense may still rely on higher-dimensional interaction vertices, in which case it is subleading in a controlled UV expansion.

An early approach to UV classification was the systematic ``topological decomposition'' of short-range dimension-nine operators presented in Ref.~\cite{Bonnet:2012kh}; see also Refs.~\cite{delAguila:2012nu,Graesser:2016bha,Helo:2015fba} for related analyses, Ref.~\cite{Gargalionis:2020} for tree-level completions of $\Delta L=2$ operators up to dim-11 without vector mediators, Ref.~\cite{Esser:2026dim9} for a recent scalar/fermion analysis of dim-9 UV models in which the dim-9 contribution can dominate, Ref.~\cite{Fonseca:2016tbn} for vector representations and gauge-boson embeddings relevant to $0\nu\beta\beta$, and Ref.~\cite{Li:2026dim11} for the recent extension to dim-11.
By enumerating tree-level topologies and classifying the quantum numbers of potential mediators, this work established a useful benchmark for the short-range dim-9 sector and serves as a reference for connecting $\onubb$ to collider-oriented models~\cite{Graf:2018ozy,Deppisch:2017ecm,Tello:2010am,Harz:2021psp,Angel:2012ug}.
However, because that language is organized after electroweak symmetry breaking, it does not by itself resolve the full spectrum of inequivalent UV completions associated with a given low-energy structure.
A purely broken-phase classification does not make the electroweak embedding unique, nor does it cleanly separate genuinely distinct electroweak completions from redundancies introduced  by contractions, permutations, or SM-like propagating fields.
This is precisely where an organization in the unbroken phase becomes valuable: it makes the role of electroweak embedding explicit and provides a clean framework in which completeness questions can be posed and answered under controlled assumptions.

In this work we carry out a systematic classification of the short-range dim-9 UV dictionary in the unbroken electroweak phase, based on the $J$-basis framework.
Starting from the dim-9 six-fermion SMEFT operator types relevant to short-range $\onubb$, we analyze the local operator space channel by channel and extract the corresponding ultraviolet mediator assignments.
For a chosen partition of external legs, this is done by diagonalizing the relevant Poincar\'e and SM gauge Casimirs.
This yields a direct resonance dictionary without manual diagram enumeration.
The resulting operator-type-level UV dictionary is complete in the unbroken phase and, in sectors with repeated fields and nontrivial permutation symmetry, less prone to omissions than a purely manual topological decomposition~\cite{Grzadkowski:2010es,Henning:2015alf,Murphy:2020rsh}.
We also provide the \texttt{Mathematica} \href{https://github.com/haolinli1991/GetUVsForType/tree/main}{code} built on our previous basis construction program \texttt{ABC4EFT} 
to automatically generate tree-level UV dictionaries for arbitrary SMEFT operator types.

Beyond the complete UV dictionary itself, the two main new results of this paper are as follows.
First, we identify the \emph{minimal ultraviolet realizations} of each operator type---the smallest set of genuinely distinct heavy mediators required after SM-like fields have been removed by field redefinitions.
The complete short-range classification yields $505$ distinct heavy-mediator combinations, of which $440$ are minimal---$12$ requiring only two and $428$ requiring three genuinely new heavy fields.
Second, we identify the massive-vector channels that survive as leading dim-9 short-range realizations.
This yields $324$ minimal UV realizations featuring vector resonances.
A cross-check against the broken-phase classification of Ref.~\cite{Bonnet:2012kh} shows that, after the strict first-generation projection, the scalar short-range sector is fully reproduced.
The paper is organized as follows.
Section~\ref{sec:eft} sets up the short-range SMEFT description and introduces the $J$-basis framework.
Section~\ref{sec:results} presents the operator-level UV dictionary, its minimal ultraviolet realizations (Section~\ref{sec:minimum-uv}).
We conclude in Section~\ref{sec:concl} with a summary and outlook.

\section{Short-range \texorpdfstring{$\onubb$}{0nuBB} in SMEFT and the \texorpdfstring{$J$}{J}-basis framework}
\label{sec:eft}

\subsection{Short-range dim-9 operator space in SMEFT}
As discussed in the Introduction, neutrinoless double beta decay can receive short-range contributions from heavy virtual fields well above the electroweak scale.
In this section we set up the EFT description of this short-range sector and specify the precise operator space that the subsequent ultraviolet classification is built upon.
Our focus is exclusively on the genuinely short-range dim-9 six-fermion sector of SMEFT: we do not revisit the long-range neutrino-exchange picture, nor the full hierarchy of $\Delta L=2$ operators at $d=5$ or $d=7$, but instead address the ultraviolet classification of the dim-9 operator space in the unbroken phase.

From the EFT point of view, the $\Delta L=2$ interactions above the electroweak scale are organized as
\begin{equation}
\mathcal{L}_{\SMEFT}^{\Delta L=2}
=
\sum_{\substack{d=5,7,9,\ldots \\ a}}
\frac{C_a^{(d)}}{\Lambda^{d-4}} \mathcal{O}_a^{(d)},
\end{equation}
where the first gauge-invariant contribution appears at $d=5$ as the Weinberg operator~\cite{Weinberg:1979sa}, while the higher odd-dimensional operators encode progressively shorter-distance realizations of $\LNV$~\cite{Babu:2001ex,deGouvea:2007xp,Bonnet:2009ej,Lehman:2014jma}.
After electroweak symmetry breaking the relevant operators are matched onto the $\Delta L=2$ low-energy EFT, and subsequently onto hadronic and nuclear degrees of freedom:
\begin{equation}
\mathcal{L}_{\SMEFT}^{\Delta L=2}
\xrightarrow[\mu \simeq m_W]{\text{EWSB and matching}}
\mathcal{L}_{\LEFT}^{\Delta L=2}
\xrightarrow[\mu \sim {\cal O}(1\text{--}2)\,\gev]{\text{hadronic matching}}
\mathcal{L}_{NN\pi ee}.
\label{eq:smeft_left_chain}
\end{equation}
The low-energy parametrization of the resulting hadronic and leptonic currents was established in Refs.~\cite{Pas:1999fc,Pas:2000vn}, while more recent EFT analyses have clarified the matching and renormalization-group structure of the short-range sector~\cite{Cirigliano:2018hja,Cirigliano:2018yza,Gonzalez:2015ady}.
In the present paper we focus on the first step in Eq.~\eqref{eq:smeft_left_chain}: the classification of gauge-invariant SMEFT operators and of their UV mediator content.
The subsequent matchings---onto the low-energy quark--lepton EFT and ultimately onto hadronic and nuclear degrees of freedom---are required for phenomenological predictions but are not rederived here; the nuclear matrix elements and isotope-dependent half-life formulas are therefore treated as downstream inputs.
It is worth noting why the dim-9 sector plays a distinguished role in this hierarchy.
At $d=5$ the Weinberg operator generates the standard light-neutrino-exchange (``long-range'') mechanism~\cite{Weinberg:1979sa}.
Higher-dimensional $\Delta L=2$ operators can also induce nonlocal or mixed realizations once electroweak symmetry is broken~\cite{Babu:2001ex,deGouvea:2007xp}.
By contrast, the first local six-fermion operators that encode genuinely short-range quark-level contributions arise systematically at $d=9$~\cite{Graesser:2016bha,Cirigliano:2018hja,Prezeau:2003xn}, making this the natural entry point for a complete ultraviolet classification of the short-range sector.
Here ``genuinely'' distinguishes the pure six-fermion operators that are irreducible at $d=9$ from effective six-fermion vertices generated by lower-dimensional $\Delta L=2$ operators---such as the dim-5 Weinberg operator or dim-7 operators involving Higgs fields---through Higgs vacuum expectation value insertions, whose Wilson coefficients are already fixed by the lower-dimensional couplings~\cite{Babu:2001ex,deGouvea:2007xp}.

In the present work we restrict attention to the SMEFT description in the unbroken electroweak phase with
\[
\Delta L = 2, \qquad \Delta B = 0,
\]
and to the genuinely short-range part of the $\onubb$ amplitude.
At low energies this sector is represented by local six-fermion operators of schematic form
\begin{equation}
\mathcal{O}^{(9)}_{\LEFT} \sim \bar u \bar u d d \bar e \bar e,
\label{eq:left_dim9_0nubb}
\end{equation}
Since the corresponding local amplitude is suppressed only by $\Lambda^{-5}$, short-range $\onubb$ provides a particularly direct bridge between low-energy searches and heavy mediator physics near the TeV scale.
In the unbroken phase, however, the structure is more than this single quark-level operator might suggest.
This variety is already visible at the operator-type level.
Restricting to the pure six-fermion sector relevant for genuine short-range dim-9 mechanisms, the SMEFT operator types are
\begin{align}
&d_{\mathbb{C}}^{\dagger 2} L^{\dagger 2} Q^{\dagger 2}, \qquad
L^{\dagger 2} Q^{2} u_{\mathbb{C}}^{2}, \qquad
d_{\mathbb{C}}^\dagger L^{\dagger 2} Q Q^\dagger u_{\mathbb{C}}, \nonumber\\
&d_{\mathbb{C}}^{\dagger 2} e_{\mathbb{C}} L^\dagger Q^\dagger u_{\mathbb{C}}, \qquad
d_{\mathbb{C}}^\dagger e_{\mathbb{C}} L^\dagger Q u_{\mathbb{C}}^{2}, \qquad
d_{\mathbb{C}}^{\dagger 2} e_{\mathbb{C}}^{2} u_{\mathbb{C}}^{2}.
\label{eq:sr6fclasses}
\end{align}
After electroweak symmetry breaking, these classes project onto the familiar quark-level chirality structures used in the broken-phase analyses of Refs.~\cite{Babu:2001ex,Bonnet:2012kh,Graesser:2016bha,Helo:2015fba}.
At the operator-type level; they cover the genuine dim-9 short-range six-fermion sector in the unbroken phase.
For the present paper, the genuine short-range dim-9 sector is therefore defined in terms of this gauge-invariant SMEFT operator space rather than a single broken-phase quark operator.
Once Lorentz, weak-isospin, and color contractions are fully resolved, the six operator types in Eq.~\eqref{eq:sr6fclasses} generate a total of $48$ independent dim-9 operators; this counting is consistent with the general SMEFT basis analyses of Refs.~\cite{Li:2020xlh,Liao:2020jmk}.

The ultraviolet classification in the unbroken-phase dim-9 sector is considerably more intricate than at lower dimensions for three related reasons.
First, at the operator-type level, the six-fermion external states generically involve repeated quark or lepton fields, so Fermi statistics restricts the allowed Lorentz, $SU(3)_C$, and $SU(2)_L$ contractions.
Second, at the gauge-embedding level, a single broken-phase chirality structure can descend from several inequivalent $\SU{2}_L$ or color contractions, each associated with a different set of ultraviolet mediators.
Third, at the UV-interpretation level, not every distinct mediator assignment represents a genuinely new heavy-field configuration: an SM-like channel can sometimes be traded for an off-shell SM insertion and absorbed by a field redefinition.

It is important to emphasize that the operator types in Eq.~\eqref{eq:sr6fclasses} do not yet coincide with the full operator basis.
Once identical fermion multiplets appear repeatedly, Fermi statistics correlates Lorentz, $\SU{2}_L$, color, and flavor structures, so that only a subset of formally admissible contractions survives as independent operators, which is what we called the $F$-basis in the framework of Young Tensor Method~\cite{Li:2020gnx,Li:2020xlh}.
These constraints are incorporated in the basis construction used below and will later play an important role in distinguishing genuinely distinct ultraviolet  channels.%
\footnote{Throughout this paper we keep the notation $u_{\mathbb C}$, $d_{\mathbb C}$, and $e_{\mathbb C}$ for the charge-conjugated right-handed Weyl fields; in the more standard SMEFT notation these correspond to $u^c$, $d^c$, and $e^c$.}

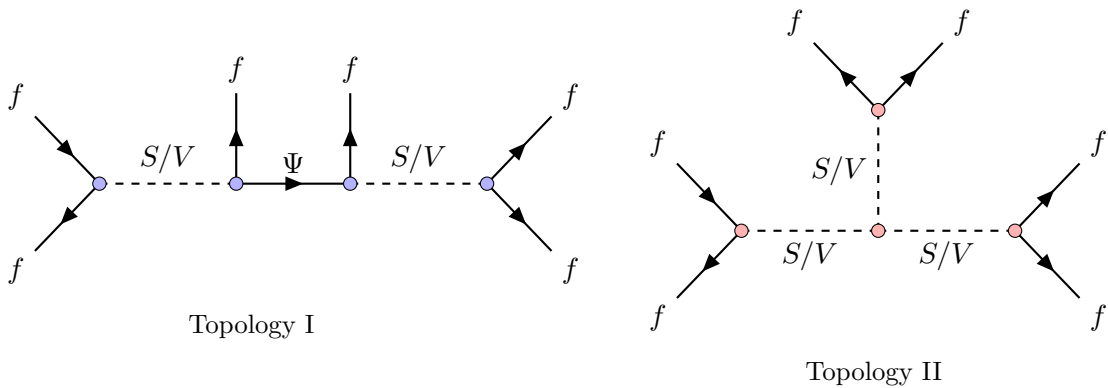
\begin{figure}[htbp]
    \centering
    \begin{minipage}{0.45\textwidth}
        \centering
        \begin{tikzpicture}
            \begin{feynman}
                \vertex (v1);
                \vertex [above left=1.2cm of v1] (i1) {$f$};
                \vertex [below left=1.2cm of v1] (i2) {$f$};
                \vertex [right=1.8cm of v1] (v2);
                \vertex [right=1.5cm of v2] (v3);
                \vertex [right=1.8cm of v3] (v4);
                \vertex [above right=1.2cm of v4] (o1) {$f$};
                \vertex [below right=1.2cm of v4] (o2) {$f$};
                \vertex [above=1.2cm of v2] (f1) {$f$};
                \vertex [above=1.2cm of v3] (f2) {$f$};

                \diagram* {
                (i1) -- [fermion, thick] (v1) -- [fermion, thick] (i2),
                (v1) -- [scalar, thick, edge label=$S/V$] (v2),
                (v2) -- [fermion, thick] (f1),
                (v2) -- [fermion, thick, edge label=$\Psi$] (v3),
                (v3) -- [fermion, thick] (f2),
                (v3) -- [scalar, thick, edge label=$S/V$] (v4),
                (v4) -- [fermion, thick] (o1),
                (v4) -- [fermion, thick] (o2),
                };
                \draw[fill=blue!30] (v1) circle (2.5pt);
                \draw[fill=blue!30] (v2) circle (2.5pt);
                \draw[fill=blue!30] (v3) circle (2.5pt);
                \draw[fill=blue!30] (v4) circle (2.5pt);
            \end{feynman}
        \end{tikzpicture}
        \vspace{0.2cm}
        {\small Topology I }
    \end{minipage}\hfill
    \begin{minipage}{0.45\textwidth}
        \centering
        \begin{tikzpicture}
            \begin{feynman}
                \vertex (v1);
                \vertex [above left=1.2cm of v1] (i1) {$f$};
                \vertex [below left=1.2cm of v1] (i2) {$f$};
                \vertex [right=1.8cm of v1] (v2);
                \vertex [right=1.8cm of v2] (v4);
                \vertex [above=1.6cm of v2] (v3);
                \vertex [above right=1.2cm of v4] (o1) {$f$};
                \vertex [below right=1.2cm of v4] (o2) {$f$};
                \vertex [above left=1.2cm of v3] (f1) {$f$};
                \vertex [above right=1.2cm of v3] (f2) {$f$};

                \diagram* {
                (i1) -- [fermion, thick] (v1) -- [fermion, thick] (i2),
                (v1) -- [scalar, thick, edge label'=$S/V$] (v2),
                (v2) -- [scalar, thick, edge label=$S/V$] (v3),
                (v2) -- [scalar, thick, edge label'=$S/V$] (v4),
                (v3) -- [fermion, thick] (f1),
                (v3) -- [fermion, thick] (f2),
                (v4) -- [fermion, thick] (o1),
                (v4) -- [fermion, thick] (o2),
                };
                \draw[fill=red!30] (v1) circle (2.5pt);
                \draw[fill=red!30] (v2) circle (2.5pt);
                \draw[fill=red!30] (v3) circle (2.5pt);
                \draw[fill=red!30] (v4) circle (2.5pt);
            \end{feynman}
        \end{tikzpicture}
        \vspace{0.2cm}
        {\small Topology II }
    \end{minipage}
    \caption{The two generic three-propagator tree topologies relevant for genuine short-range $d=9$ $\onubb$ operators. External lines ($f$) are SM fermions, while the internal lines denote heavy bosonic or fermionic mediators. Blue and red dots indicate gauge-invariant interaction vertices in Topology~I and Topology~II, respectively. In the present paper these topologies are embedded into an analysis performed in the unbroken electroweak phase, so that inequivalent $\SU{2}_L$ and color channels are kept separate before symmetry breaking.}
    \label{fig:topologies}
\end{figure}

Broken-phase topological decompositions, as explored in Ref.~\cite{Bonnet:2012kh} (see also Ref.~\cite{Bonnet:2012kz} for the $d=5$ case), offer a useful guide to the generic tree structures relevant for short-range $\onubb$.
For genuine dim-9 six-fermion mechanisms, there are two kinds of three-propagator topologies shown in Figure~\ref{fig:topologies}.
Topology~I contains a boson--fermion--boson chain, whereas Topology~II contains three bosonic propagators.
What these topologies do not determine is which electroweak representations flow in the internal lines; this is precisely why the SMEFT operator-level organization introduced below is needed.
These topologies provide the diagrammatic matching templates for the short-range sector, which provides the partition information for the J-basis analysis for the SMEFT operators of the type listed in Eq.~\eqref{eq:sr6fclasses} as will be discussed in the next section.

To summarize: the ultraviolet classification pursued in the rest of 
this paper consists of identifying, for each independent dim-9 SMEFT 
operator types in of Eq.~\eqref{eq:sr6fclasses}, 
all sets of heavy mediators that can generate it through a connected 
tree-level graph with renormalizable vertices.
The precise criteria---including tree closure and leading-power 
constraints---are specified in Section~\ref{sec:channel-to-uv}.


\subsection{\texorpdfstring{$J$}{J}-basis organization of the local operator space}
\label{sec:jbasis-method}

The central technical ingredient of the present work is the $J$-basis organization of the SMEFT operator space.
Its role here is to keep the ultraviolet classification in the unbroken phase, where the full electroweak structure of the operator space is retained.
This is the essential difference from a broken-phase or LEFT-oriented description: in the latter, distinct SMEFT completions may already be merged into the same low-energy quark-level structure, whereas in the SMEFT $J$-basis they remain separated as inequivalent weak-isospin channels.

The method used here is part of a broader on-shell and group-theoretic program developed in earlier works~\cite{Li:2020gnx,Li:2020xlh}.
The operator--amplitude correspondence established in Ref.~\cite{Ma:2019gtx,Durieux:2019eor,Henning:2019enq} provides the foundation for turning the local-operator problem into a finite-dimensional algebraic one.
Systematic constructions of complete higher-dimensional SMEFT bases then supplied the independent and complete operator spaces on which the method acts~\cite{Li:2020gnx,Li:2020xlh,Li:2022tec,Henning:2015alf,Murphy:2020rsh}.
On this foundation, the generalized partial-wave framework for arbitrary multi-particle channels~\cite{Jiang:2020rwz,Shu:2021qlr} and the bottom-up UV-resonance program~\cite{Li:2022abx,Li:2023dim8uv} lead directly to the $J$-basis interpretation used here.
The present paper applies this logic to the dim-9 $\Delta L=2$ six-fermion sector relevant for short-range $\onubb$.

In a zoo of SMEFT operator bases in Refs.~\cite{Li:2020gnx,Li:2022tec}, different bases emphasize different aspects of the EFT: the $Y$-basis ensures algebraic independence, the $P$-basis makes the flavor and permutation structure of repeated fields manifest, while the $J$-basis reorganizes the independent operator space according to conserved channel quantum numbers and thereby prepares the ultraviolet interpretation.
To be more specific, it reorganizes the EFT operators within a complete and independent basis such that the corresponding local on-shell amplitudes carry definite quantum numbers for different groups of particles, which we refer to as channels.
Owing to this correspondence, these quantum numbers may be interpreted as those of UV particles appearing in diagrammatic matching. In this way, the UV interpretation is reduced to a problem in representation theory.

To define the $J$-basis, one must first specify all channels, which is equivalent to choosing a partition.
By virtue of the amplitude--operator correspondence, the fields appearing in an operator are identified with the external particles in the local on-shell amplitude generated by that operator.
Mathematically, a partition is a collection of subsets, $\mathcal{P}=\{I_1,\ldots,I_m\}$, where each $I_i\subset \{1,\ldots,N\}$ and where $1,\ldots,N$ label the external particles.
These subsets are required to satisfy two conditions: first, $\bigcup_{r=1}^{m} I_r = \{1,\ldots,N\}$; second, any two subsets must be either disjoint or nested, i.e.\ $I_r\cap I_s=\varnothing$, $I_r\subset I_s$, or $I_s\subset I_r$.
This definition extends the familiar channel decomposition of ordinary $2\to2$ partial waves to general tree-level multi-particle factorizations.
Diagrammatically, the subset notation $\mathcal{P}$ is equivalent to a tree diagram with fixed external particles in a top-down matching computation.
It is precisely this correspondence between partitions and tree diagrams that later enables the quantum numbers associated with a channel $I_i$ to be interpreted as those of the corresponding mediator in the tree diagram.

For the short-range six-fermion sector considered in Eq.~\eqref{eq:sr6fclasses}, the relevant partitions are those compatible with Topology~I and Topology~II in Fig.~\ref{fig:topologies}; together they exhaust the tree-level UV realization of interest.
Representative examples are shown in Table~\ref{tab:partition_examples}.
The leftmost column displays the tree diagrams obtained by fixing all external legs in the topologies of Fig.~\ref{fig:topologies}, the middle column gives our partition notation explicitly as sets of subsets, and the rightmost column shows the notation adopted in Ref.~\cite{Bonnet:2012kh}.

\begin{table}[t]
\centering
\renewcommand{\arraystretch}{1.12}
\small
\begin{tabular}{@{}>{\centering\arraybackslash}m{0.28\textwidth}@{\hspace{3.5em}}>{\raggedright\arraybackslash}m{0.36\textwidth}@{\hspace{0.5em}}>{\raggedright\arraybackslash}m{0.24\textwidth}@{}}
\toprule
Topology  & Partition & Broken-phase decomposition \\
\midrule
\parbox{0.27\textwidth}{\centering
\scalebox{0.40}{%
\begin{tikzpicture}[baseline=(current bounding box.center)]
  \begin{feynman}
      \vertex (v1);
      \vertex [above left=1.0cm of v1] (i1) {$Q$};
      \vertex [below left=1.0cm of v1] (i2) {$u_{\mathbb C}$};
      \vertex [right=1.45cm of v1] (v2);
      \vertex [right=1.15cm of v2] (v3);
      \vertex [right=1.45cm of v3] (v4);
      \vertex [above right=1.0cm of v4] (o1) {$L^\dagger$};
      \vertex [below right=1.0cm of v4] (o2) {$L^\dagger$};
      \vertex [above=1.0cm of v2] (f1) {$Q^\dagger$};
      \vertex [above=1.0cm of v3] (f2) {$d_{\mathbb C}^\dagger$};
      \diagram* {
      (i1) -- [fermion, thick] (v1) -- [fermion, thick] (i2),
      (v1) -- [scalar, thick, edge label=$S$] (v2),
      (v2) -- [fermion, thick] (f1),
      (v2) -- [fermion, thick, edge label=$\Psi$] (v3),
      (v3) -- [fermion, thick] (f2),
      (v3) -- [scalar, thick, edge label=$S$] (v4),
      (v4) -- [fermion, thick] (o1),
      (v4) -- [fermion, thick] (o2),
      };
      \draw[fill=blue!30] (v1) circle (2.3pt);
      \draw[fill=blue!30] (v2) circle (2.3pt);
      \draw[fill=blue!30] (v3) circle (2.3pt);
      \draw[fill=blue!30] (v4) circle (2.3pt);
  \end{feynman}
\end{tikzpicture}}\\[2pt]
{\footnotesize Topology~I (a)}} &
\makecell[l]{$\mathcal P_{\rm I}^{(a)}=\{I_1,I_2,I_3\}$,\\[0.4ex]
$I_1=(Q\,u_{\mathbb C})$,\\[0.4ex]
$I_2=(Q\,u_{\mathbb C})\,Q^\dagger = d_{\mathbb C}^\dagger(L^\dagger L^\dagger)$,\\[0.4ex]
$I_3=(L^\dagger L^\dagger)$}
&
\makecell[l]{\#1-ii-a:\\[0.4ex]
$(Q\,u_{\mathbb C})\,(Q^\dagger)\,(d_{\mathbb C}^\dagger)\,(L^\dagger L^\dagger)$}
\\[1.0ex]
\midrule
\parbox{0.27\textwidth}{\centering
\scalebox{0.40}{%
\begin{tikzpicture}[baseline=(current bounding box.center)]
  \begin{feynman}
      \vertex (v1);
      \vertex [above left=1.0cm of v1] (i1) {$Q$};
      \vertex [below left=1.0cm of v1] (i2) {$u_{\mathbb C}$};
      \vertex [right=1.45cm of v1] (v2);
      \vertex [right=1.15cm of v2] (v3);
      \vertex [right=1.45cm of v3] (v4);
      \vertex [above right=1.0cm of v4] (o1) {$L^\dagger$};
      \vertex [below right=1.0cm of v4] (o2) {$Q^\dagger$};
      \vertex [above=1.0cm of v2] (f1) {$d_{\mathbb C}^\dagger$};
      \vertex [above=1.0cm of v3] (f2) {$L^\dagger$};
      \diagram* {
      (i1) -- [fermion, thick] (v1) -- [fermion, thick] (i2),
      (v1) -- [scalar, thick, edge label=$S$] (v2),
      (v2) -- [fermion, thick] (f1),
      (v2) -- [fermion, thick, edge label=$\Psi$] (v3),
      (v3) -- [fermion, thick] (f2),
      (v3) -- [scalar, thick, edge label=$S$] (v4),
      (v4) -- [fermion, thick] (o1),
      (v4) -- [fermion, thick] (o2),
      };
      \draw[fill=blue!30] (v1) circle (2.3pt);
      \draw[fill=blue!30] (v2) circle (2.3pt);
      \draw[fill=blue!30] (v3) circle (2.3pt);
      \draw[fill=blue!30] (v4) circle (2.3pt);
  \end{feynman}
\end{tikzpicture}}\\[2pt]
{\footnotesize Topology~I (b)}} &
\makecell[l]{$\mathcal P_{\rm I}^{(b)}=\{I_1,I_2,I_3\}$,\\[0.4ex]
$I_1=(Q\,u_{\mathbb C})$,\\[0.4ex]
$I_2=(Q\,u_{\mathbb C})\,d_{\mathbb C}^\dagger = L^\dagger(L^\dagger Q^\dagger)$,\\[0.4ex]
$I_3=(L^\dagger Q^\dagger)$}
&
\makecell[l]{\#2-i-a:\\[0.4ex]
$(Q\,u_{\mathbb C})\,(d_{\mathbb C}^\dagger)\,(L^\dagger)\,(L^\dagger Q^\dagger)$}
\\[1.0ex]
\midrule
\parbox{0.27\textwidth}{\centering
\scalebox{0.40}{%
\begin{tikzpicture}[baseline=(current bounding box.center)]
  \begin{feynman}
      \vertex (v1);
      \vertex [above left=1.0cm of v1] (i1) {$Q$};
      \vertex [below left=1.0cm of v1] (i2) {$u_{\mathbb C}$};
      \vertex [right=1.55cm of v1] (v2);
      \vertex [right=1.55cm of v2] (v4);
      \vertex [above=1.35cm of v2] (v3);
      \vertex [above right=1.0cm of v4] (o1) {$Q$};
      \vertex [below right=1.0cm of v4] (o2) {$u_{\mathbb C}$};
      \vertex [above left=0.9cm of v3] (f1) {$L^\dagger$};
      \vertex [above right=0.9cm of v3] (f2) {$L^\dagger$};
      \diagram* {
      (i1) -- [fermion, thick] (v1) -- [fermion, thick] (i2),
      (v1) -- [scalar, thick, edge label'=$S$] (v2),
      (v2) -- [scalar, thick, edge label=$S$] (v3),
      (v2) -- [scalar, thick, edge label'=$S$] (v4),
      (v3) -- [fermion, thick] (f1),
      (v3) -- [fermion, thick] (f2),
      (v4) -- [fermion, thick] (o1),
      (v4) -- [fermion, thick] (o2),
      };
      \draw[fill=red!30] (v1) circle (2.3pt);
      \draw[fill=red!30] (v2) circle (2.3pt);
      \draw[fill=red!30] (v3) circle (2.3pt);
      \draw[fill=red!30] (v4) circle (2.3pt);
  \end{feynman}
\end{tikzpicture}}\\[2pt]
{\footnotesize Topology~II (a)}} &
\makecell[l]{$\mathcal P_{\rm II}^{(a)}=\{I_1,I_2,I_3\}$,\\[0.4ex]
$I_1=(Q\,u_{\mathbb C})_{L}$, $I_2=(L^\dagger L^\dagger)$,\\[0.4ex]
$I_3=(Q\,u_{\mathbb C})_{R}$}
&
\makecell[l]{\#1:\\[0.4ex]
$(Q\,u_{\mathbb C})\,(L^\dagger L^\dagger)\,(Q\,u_{\mathbb C})$}
\\[1.0ex]
\midrule
\parbox{0.27\textwidth}{\centering
\scalebox{0.40}{%
\begin{tikzpicture}[baseline=(current bounding box.center)]
  \begin{feynman}
      \vertex (v1);
      \vertex [above left=1.0cm of v1] (i1) {$L^\dagger$};
      \vertex [below left=1.0cm of v1] (i2) {$d_{\mathbb C}^\dagger$};
      \vertex [right=1.55cm of v1] (v2);
      \vertex [right=1.55cm of v2] (v4);
      \vertex [above=1.35cm of v2] (v3);
      \vertex [above right=1.0cm of v4] (o1) {$Q$};
      \vertex [below right=1.0cm of v4] (o2) {$u_{\mathbb C}$};
      \vertex [above left=0.9cm of v3] (f1) {$L^\dagger$};
      \vertex [above right=0.9cm of v3] (f2) {$Q^\dagger$};
      \diagram* {
      (i1) -- [fermion, thick] (v1) -- [fermion, thick] (i2),
      (v1) -- [scalar, thick, edge label'=$S$] (v2),
      (v2) -- [scalar, thick, edge label=$S$] (v3),
      (v2) -- [scalar, thick, edge label'=$S$] (v4),
      (v3) -- [fermion, thick] (f1),
      (v3) -- [fermion, thick] (f2),
      (v4) -- [fermion, thick] (o1),
      (v4) -- [fermion, thick] (o2),
      };
      \draw[fill=red!30] (v1) circle (2.3pt);
      \draw[fill=red!30] (v2) circle (2.3pt);
      \draw[fill=red!30] (v3) circle (2.3pt);
      \draw[fill=red!30] (v4) circle (2.3pt);
  \end{feynman}
\end{tikzpicture}}\\[2pt]
{\footnotesize Topology~II (b)}} &
\makecell[l]{$\mathcal P_{\rm II}^{(b)}=\{I_1,I_2,I_3\}$,\\[0.4ex]
$I_1=(L^\dagger d_{\mathbb C}^\dagger)$, $I_2=(L^\dagger Q^\dagger)$,\\[0.4ex]
$I_3=(Q\,u_{\mathbb C})$}
&
\makecell[l]{\#2:\\[0.4ex]
$(L^\dagger d_{\mathbb C}^\dagger)\,(L^\dagger Q^\dagger)\,(Q\,u_{\mathbb C})$}
\\[1.0ex]
\bottomrule
\end{tabular}
\caption{Representative channel assignments for the two six-fermion tree topologies. Each row shows one explicit external-fermion assignment together with a simplified notation $\mathcal P=\{I_1,I_2,I_3\}$, in which the three $I_r$ are identified with the three propagator channels of the tree and written directly in terms of the corresponding fermion currents; the last column records the associated broken-phase fermion grouping. Repeated identical endpoint currents are distinguished by the subscripts $L$ and $R$.}
\label{tab:partition_examples}
\end{table}

Through the operator--amplitude correspondence, the same partition can be applied to the associated on-shell contact amplitude. For a fixed partition $\mathcal P$, the relevant statement is the correspondence between the finite-dimensional local operator space and the associated local amplitude space,
\begin{equation}
\mathcal{O}_a
\;\longleftrightarrow\;
\mathcal{M}_a(\{\Psi\}_{\mathcal{P}}).
\label{eq:operator_amplitude_map}
\end{equation}
Here $\mathcal{M}_a(\{\Psi\}_{\mathcal{P}})$ denotes the local on-shell amplitude corresponding to $\mathcal{O}_a$ in the chosen channel partition. What we need from the generalized partial-wave framework~\cite{Jiang:2020rwz,Shu:2021qlr} is therefore not a dynamical scattering expansion, but the channel-wise Casimir construction on the amplitude side. Since the operator--amplitude map preserves the relevant symmetry representations, the spin and gauge quantum numbers extracted there can be transferred back directly to the finite-dimensional operator space $\mathcal{V}_{\rm SR}^{(9)}$.

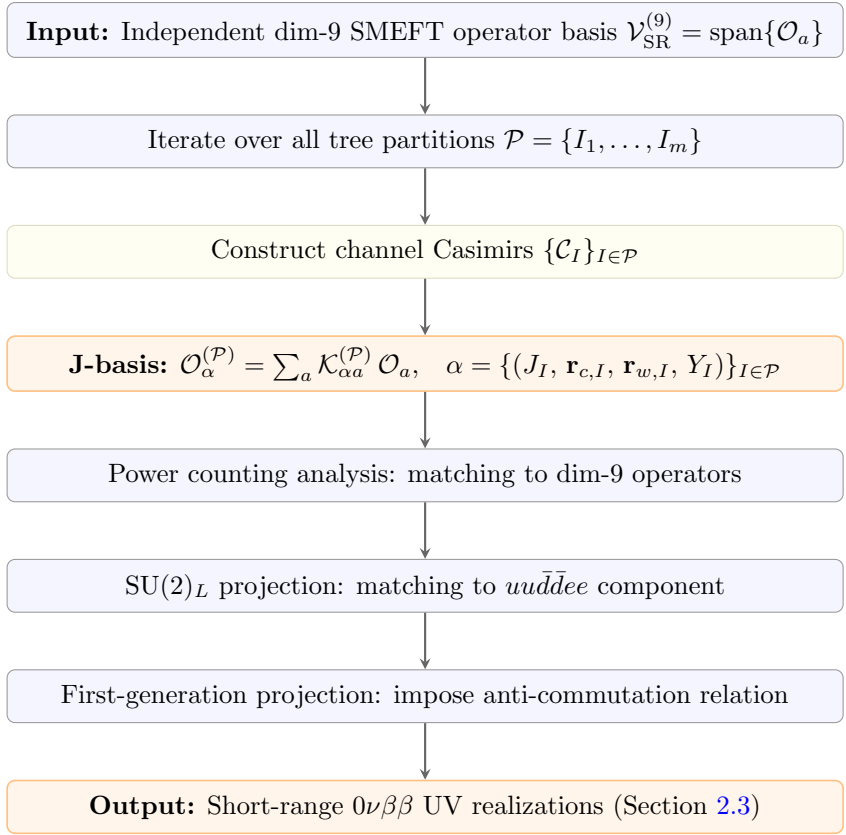
\begin{figure}[tp]
\centering
\begin{tikzpicture}[
  node distance=0.75cm,
  box/.style={rectangle, draw, rounded corners=3pt,
              minimum width=11cm, minimum height=0.7cm,
              align=center, font=\small,
              fill=blue!4, draw=black!40},
  prepbox/.style={box, fill=yellow!6, draw=yellow!50!black!30},
  corebox/.style={box, fill=orange!8, draw=orange!50, line width=0.6pt},
  arrow/.style={->, thick, >=stealth, color=black!60}
]
\node[box] (basis)
  {\textbf{Input:} Independent dim-9 SMEFT operator basis
   $\mathcal{V}_{\rm SR}^{(9)}=\mathrm{span}\{\mathcal{O}_a\}$};
\node[box, below=of basis] (partition)
  {Iterate over all tree partitions
   $\mathcal{P}=\{I_1,\ldots,I_m\}$};
\node[prepbox, below=of partition] (casimirs)
  {Construct channel Casimirs
   $\{\mathcal{C}_{I}\}_{I\in\mathcal{P}}$};
\node[corebox, below=of casimirs] (jbasis)
  {\textbf{J-basis:}
   $\mathcal{O}_\alpha^{(\mathcal{P})}=\sum_a \mathcal{K}_{\alpha a}^{(\mathcal{P})}\,\mathcal{O}_a$,\quad
   $\alpha=\{(J_I,\,\mathbf{r}_{c,I},\,\mathbf{r}_{w,I},\,Y_I)\}_{I\in\mathcal{P}}$};
\node[box, below=of jbasis] (power)
  {Power counting analysis:
   matching to dim-9 operators};
\node[box, below=of power] (weak)
  {$\SU{2}_L$ projection:
   matching to $uu\bar{d}\bar{d}ee$ component};
\node[box, below=of weak] (gen1)
  {First-generation projection:
   impose anti-commutation relation};
\node[corebox, below=of gen1] (uv)
  {\textbf{Output:} Short-range $\onubb$ UV realizations
   (Section~\ref{sec:channel-to-uv})};
\draw[arrow] (basis)      -- (partition);
\draw[arrow] (partition)   -- (casimirs);
\draw[arrow] (casimirs)    -- (jbasis);
\draw[arrow] (jbasis)      -- (power);
\draw[arrow] (power)       -- (weak);
\draw[arrow] (weak)        -- (gen1);
\draw[arrow] (gen1)        -- (uv);
\end{tikzpicture}
\caption{Schematic workflow of the $J$-basis method.
  Starting from an independent operator basis, one iterates over all tree partitions,
  represents the corresponding commuting Casimir operators on the local amplitude space,
  and finds their common eigenbasis to obtain the $J$-basis operators
  whose eigenvalues encode the admissible UV mediator quantum numbers.
  The yellow box indicates the construction of the commuting
  Casimir family; the orange boxes highlight the core diagonalization step
  and its output. The final three boxes indicate the additional physical filters
  used in the strict short-range $\onubb$ classification.}
\label{fig:jbasis_flowchart}
\end{figure}

For each channel $I$ we define the partial kinematic generators by summing over the legs belonging to that channel, $P_I^\mu=\sum_{i\in I}p_i^\mu$ and $M_I^{\mu\nu}=\sum_{i\in I}M_i^{\mu\nu}$, where $M_i^{\mu\nu}$ denotes the Lorentz generator acting on the $i$-th external leg in the on-shell representation of the contact amplitude.
From these generators one constructs the partial Pauli--Lubanski vector
\begin{equation}
W_I^\mu
=
\frac12 \epsilon^{\mu\nu\rho\sigma}
P_{I,\nu}\,
M_{I,\rho\sigma},
\qquad
W_I^2
=
W_{I,\mu}W_I^\mu,
\label{eq:partial_PL}
\end{equation}
which measures the Lorentz spin carried in the channel.
Similarly, for each SM gauge factor $G$ one defines the partial generators
\begin{equation}
T_I^{A,(G)}=\sum_{i\in I}T_i^{A,(G)},
\qquad
C_{2,I}^{(G)}=T_I^{A,(G)}T_I^{A,(G)}.
\label{eq:partial_gauge_casimir}
\end{equation}
More generally, we denote by $C_{n,I}^{(G)}$ the independent Casimir operators required to distinguish the relevant irreducible representations in channel $I$; for $\SU{3}_C$ this includes the cubic Casimir whenever distinct color irreps share the same quadratic eigenvalue.
For the abelian factor $U(1)_Y$, there is no nontrivial Casimir. Instead, the hypercharge flowing in channel $I$ is simply the additive label
\(
Y_I=\sum_{i\in I}Y_i
\),
which is fixed once the external legs in that channel are specified.
After choosing a basis for the local amplitude space $\mathcal{M}_{\rm loc}^{(9)}(\{\Psi\}_{\mathcal{P}})$, the operators $W_I^2$ and $C_{n,I}^{(G)}$ are represented by finite-dimensional matrices. For compatible channels $I,J\in\mathcal{P}$ (i.e.\ disjoint or nested), the partial Lorentz and gauge Casimirs commute on this amplitude space~\cite{Li:2022abx}:
\begin{equation}
\bigl[W_I^2,W_J^2\bigr]
=
\bigl[W_I^2,C_{m,J}^{(G')}\bigr]
=
\bigl[C_{n,I}^{(G)},C_{m,J}^{(G')}\bigr]
=0.
\label{eq:compatible_commutators}
\end{equation}
This is the algebraic reflection of the fact that a tree-level intermediate state carries simultaneously definite spin and gauge quantum numbers in every channel. The set $\{W_I^2,\,C_{n,I}^{(G)}\}_{I\in\mathcal{P},\,G}$ therefore defines a commuting family of matrices on $\mathcal{M}_{\rm loc}^{(9)}(\{\Psi\}_{\mathcal{P}})$, which can be diagonalized simultaneously. The corresponding common eigenamplitudes satisfy
\begin{align}
W_I^2\,\mathcal{M}_{\alpha}^{(\mathcal{P})}
&=
-s_I\,J_I(J_I+1)\,
\mathcal{M}_{\alpha}^{(\mathcal{P})},
\label{eq:Jbasis_W2}
\\
C_{n,I}^{(G)}\,\mathcal{M}_{\alpha}^{(\mathcal{P})}
&=
C_n^{(G)}\!\bigl(\mathbf{r}_I^{(G)}\bigr)\,
\mathcal{M}_{\alpha}^{(\mathcal{P})},
\qquad
G=\SU{3}_C,\,\SU{2}_L,
\label{eq:Jbasis_Cn}
\end{align}
for every channel $I\in\mathcal{P}$.
Transporting this common eigenbasis back through the operator--amplitude correspondence defines the $J$-basis on the operator side. Expanded on the original operator basis, it takes the form~\cite{Li:2022abx,Li:2023dim8uv}
\begin{equation}
\mathcal{O}_{\alpha}^{(\mathcal{P})}
=
\sum_a \mathcal{K}_{\alpha a}^{(\mathcal{P})}\,\mathcal{O}_a,
\label{eq:Jbasis_def}
\end{equation}
where $\mathcal{K}^{(\mathcal{P})}$ is the change-of-basis matrix induced by the simultaneous diagonalization on the amplitude side.
The sequence of steps leading from the independent operator space to the channel-wise UV data is summarized in Figure~\ref{fig:jbasis_flowchart}.

The common eigenbasis obtained in this way has a direct UV interpretation. For each channel $I$, the eigenvalues determine the full one-particle label $(J_I,\,\mathbf{r}_{c,I},\,\mathbf{r}_{w,I},\,Y_I)$: the partial Pauli--Lubanski Casimir fixes the spin, the non-abelian gauge Casimirs fix the $\SU{3}_C$ and $\SU{2}_L$ representations, and the hypercharge is fixed additively from the external legs. In this sense, the $J$-basis converts the search for admissible tree-level mediator quantum numbers into a channel-wise eigenvalue problem on a finite-dimensional local operator space.

The remaining step is to determine which sets of algebraically admissible channel labels can actually be assembled into genuine ultraviolet realizations with connected tree structure and the correct power counting. This is the task of the next subsection.

\subsection{From channel quantum numbers to ultraviolet realizations}
\label{sec:channel-to-uv}

The output of the $J$-basis diagonalization for a given partition $\mathcal{P}$ may be summarized as
\begin{align}
\mathcal{O}_{\alpha}^{(\mathcal{P})}
\quad \longmapsto \quad
\Bigl\{J_I,\, \mathbf{r}_{c,I},\, \mathbf{r}_{w,I},\, Y_I\Bigr\}_{I\in\mathcal{P}},
\label{eq:channel_correspondence}
\end{align}
which assigns a candidate one-particle label to every channel of the chosen tree partition. For the dim-9 short-range $\onubb$ problem, however, this algebraic output is only the starting point. The reason is physical: the classification is carried out in electroweak-complete SMEFT in the unbroken phase, whereas the observable $\onubb$ process is described after electroweak symmetry breaking and picks out the specific broken-phase component $uudd\,ee$. Before a channel assignment is retained in the physical catalogue, we therefore impose three additional filters: leading-power counting, projection to the physical weak component, and the strict first-generation constraints.

\paragraph*{Leading-power counting.}
The first filter keeps only channels that can contribute already at order $\Lambda^{-5}$ through gauge-invariant renormalizable interactions. For scalar and fermion mediators this introduces no additional subtlety: once the channel quantum numbers admit the required renormalizable vertices, the corresponding tree contributes directly to the dim-9 six-fermion sector. Massive vectors are different. A vector channel that is algebraically allowed by the $J$-basis need not be a leading dim-9 realization, because its endpoint couplings may first appear only through higher-dimensional operators. The characteristic problematic case is a same-chirality fermion--fermion--vector endpoint, for which the first nonvanishing local interaction is the dim-5 tensor-current coupling
\begin{equation}
\frac{1}{\Lambda}\,(\psi \sigma_{\mu\nu}\chi)\,V^{\mu\nu}.
\label{eq:tensor_current_vertex}
\end{equation}
Such an endpoint is one power of $\Lambda$ more suppressed than a renormalizable current coupling, so a branch that would otherwise look like a dim-9 tree is demoted to $d\ge 10$ and is not counted as a leading short-range realization. The physical question is therefore not whether vector exchange is allowed, but whether the corresponding channel can be realized by a genuinely renormalizable tree already at leading order. In the explicit dim-9 scan this subtlety affects only a small subset of vector-mediated classifications, and in every surviving case the same heavy-field content also admits an ordinary renormalizable tree realization. The role of the power-counting filter is thus to remove nonleading tensor-current routes without discarding vector multiplets that genuinely contribute already at dim~9. 

\paragraph*{Projection to the physical weak component.}
The second filter appears when one descends from the electroweak-complete SMEFT operator to the broken-phase process. An operator that is perfectly admissible in the unbroken phase need not contribute to $\onubb$, because the latter probes only the particular weak component $uu\bar{d}\bar{d}\,ee$. In this step the detailed $\SU{2}_L$ contractions matter. Weak-singlet contractions of identical doublets can project away the required state entirely: an $LL$ or $L^\dagger L^\dagger$ singlet contains no $ee$, while a $QQ$ or $Q^\dagger Q^\dagger$ singlet contains no $dd$ or $uu$. The weak structure must therefore be followed through the full completion, and only those branches with a nonvanishing $uu\bar{d}\bar{d}\,ee$ projection are retained in the physical short-range catalogue.

\paragraph*{First-generation projection.}
After the weak projection has selected the physical component, one must still impose the additional constraints that arise because the physical $\onubb$ channel involves repeated first-generation fermions. Once the external fields are fixed to $u$, $d$, and $e$, fermion-exchange antisymmetry becomes a genuine selection rule: the relevant amplitudes must respect the combined Lorentz, color, and weak symmetry properties of identical quark or lepton pairs. As a result, some UVs that are admissible for generic flavor assignments vanish identically in the first-generation limit. This last filter is therefore applied at the level of complete tree realizations rather than individual channels, because the cancellation can depend on how the repeated external lines are tied together across the full diagram.

The short-range dim-9 catalogue studied in Section~\ref{sec:results} is obtained only after all three filters have been imposed. The $J$-basis thus supplies the complete algebraic channel data, while the subsequent power-counting, weak-projection, and first-generation steps identify the subset that contributes to the physical $\onubb$ process.

\section{Ultraviolet realizations of the dim-9 operator space}
\label{sec:results}

This section first fixes the convention for handling SM-like ultraviolet fields and then summarizes the heavy-resonance content and minimal-UV classification of the short-range dim-9 operator space, including the surviving massive-vector sector.

\subsection{SM-like fields and the field-redefinition convention}

Before presenting the short-range classification, a clear interpretational convention must be established.
The underlying bottom-up EFT framework---incorporating operator-amplitude correspondence and field redefinition reductions---was systematically developed in Refs.~\cite{Li:2022tec,Li:2022abx}. In the following, we adopt only the essential conventions required for the dimension-nine short-range classification.

A subtle but essential consideration involves channels with quantum numbers—spanning spin and all gauge symmetries—that coincide with those of Standard Model (SM) fields; we define these as 'SM-like' channels.
In many cases, a heavy degree of freedom is not strictly required in such a channel for the generation of a short-range process. Instead, the target effective operator may emerge from the field redefinition reduction of a sibling operator type---one involving fewer fields but containing additional derivatives---where the SM field of the same representation mediates the interaction. Consequently, these SM-like channels are typically excluded from the count of genuine heavy degrees of freedom when identifying minimal UV realizations.

\paragraph*{A dim-9 example.}

A representative example is a Higgs-like insertion taken from the operator type $d_{\mathbb{C}}^{\dagger 2}L^{\dagger 2}Q^{\dagger 2}$.
In this Topology~II completion the bilinears
$d_{\mathbb{C}}^\dagger L^\dagger$ and $L^\dagger Q^\dagger$
are matched to the channels
$S_{20}\,(\mathbf{3},\mathbf{2},1/6)$ and
$S_{15}^\dagger\,(\bar{\mathbf{3}},\mathbf{1},1/3)$,
while the color-singlet part of
$d_{\mathbb{C}}^\dagger Q^\dagger$
carries the Higgs quantum numbers
$S_{4}\,(\mathbf{1},\mathbf{2},1/2)\sim H$.
Once the genuinely heavy lines $S_{15}^\dagger$ and $S_{20}$ are integrated out, the remaining $S_4$ channel may be reinterpreted as an off-shell $H^\dagger$ insertion rather than as an additional heavy resonance in the ultraviolet spectrum.

The second viewpoint produces an operator of the form
\begin{equation}
\mathcal O_{\rm off}
\sim
(D^2 H^\dagger)\,(d_{\mathbb{C}}^\dagger Q^\dagger)(L^\dagger L^\dagger),
\label{eq:dim9_eom_example}
\end{equation}
where only the operator type is displayed and the detailed Lorentz, color, and weak-index contractions are suppressed.
This operator is redundant in our on-shell operator basis convention. It can be removed by the Higgs field redefinition
\begin{equation}
H^\dagger \;\rightarrow\;
H^\dagger + \frac{c}{\Lambda^5}(d_{\mathbb{C}}^\dagger Q^\dagger)(L^\dagger L^\dagger),
\label{eq:dim9_field_redefinition}
\end{equation}
When applying this field redefinition to the kinetic term of the Higgs fields, it will generate the term that cancels the ${\cal O}_{\rm off}$ in Eq.~\eqref{eq:dim9_eom_example} in the Lagrangian, while the Higgs Yukawa term generates the six-fermion operators in our basis.
Schematically, this is equivalent to the replacement of $D^2 H^\dagger$ with equation of motion (EOM) of the Higgs field at the renormalizable level in the off-shell operator when truncating to dimension nine:
\begin{equation}
D^2 H^\dagger
\;\sim\;
\frac{\partial V(H)}{\partial H}
\, + \,
Y_d^\dagger\,Q^\dagger d_{\mathbb C}^\dagger
\, + \,
Y_e^\dagger\,L^\dagger e_{\mathbb C}^\dagger
\,+\,\cdots ,
\label{eq:higgs_eom_schematic}
\end{equation}
\begin{equation}
\mathcal O_{\rm off}
\;\xrightarrow{\mathrm{EOM}_{H^\dagger}}\;
 (d_{\mathbb{C}}^\dagger L^\dagger)(L^\dagger Q^\dagger)
\bigl(Y_d^\dagger\,Q^\dagger d_{\mathbb{C}}^\dagger
      + Y_e^\dagger\,L^\dagger e_{\mathbb{C}}^\dagger\bigr)
\,+\,\cdots ,
\label{eq:dim9_eom_mapping}
\end{equation}
where the ellipses again denote bosonic terms or structures outside the short-range six-fermion sector.
The $Y_d^\dagger$ term is the piece that feeds the short-range six-fermion sector relevant to $\onubb$. 
These schematic relations make explicit why one can omit the heavy Higgs-like mediator in the UV theory while still obtaining the six-fermion operator of interest.

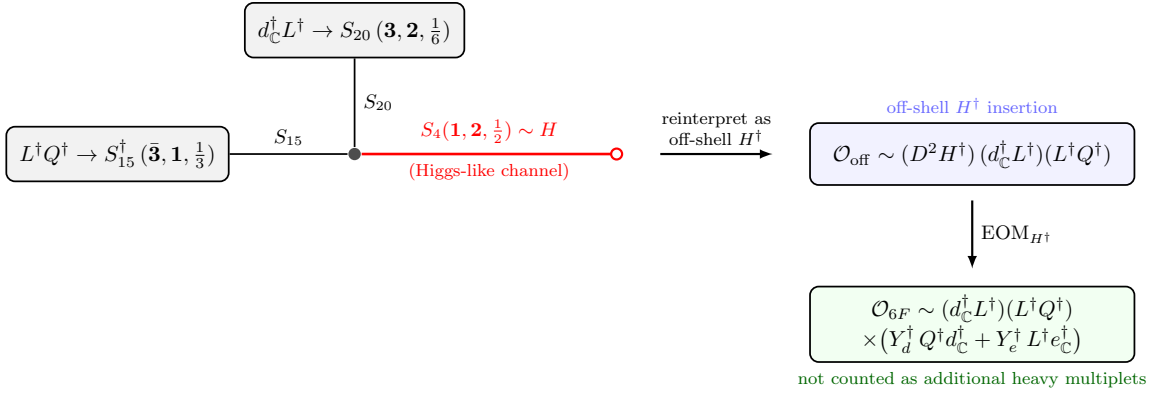
\begin{figure}[t]
\centering
\resizebox{\textwidth}{!}{%
\begin{tikzpicture}[>=latex, thick, every node/.style={font=\small}]

\node[draw, rounded corners, fill=gray!10, minimum height=0.85cm, inner sep=6pt]
      (ll) at (-3.5,0) {$L^\dagger Q^\dagger \to S_{15}^\dagger\,(\bar{\mathbf{3}},\mathbf{1},\tfrac{1}{3})$};
\node[draw, rounded corners, fill=gray!10, minimum height=0.85cm, inner sep=6pt]
      (dq) at (0.3,2.0) {$d_{\mathbb{C}}^\dagger L^\dagger \to S_{20}\,(\mathbf{3},\mathbf{2},\tfrac{1}{6})$};
\node[circle, fill=black!70, inner sep=2pt] (c) at (0.3,0) {};
\draw[thick] (ll.east) -- node[above, font=\footnotesize] {$S_{15}$} (c);
\draw[thick] (dq.south) -- node[right, font=\footnotesize] {$S_{20}$} (c);
\coordinate (hend) at (4.5,0);
\draw[thick, red, line width=1.2pt] (c) --
      node[above, font=\footnotesize, yshift=1pt] {\color{red}$S_4(\mathbf{1},\mathbf{2},\tfrac{1}{2}) \sim H$}
      node[below, font=\scriptsize, red] {(Higgs-like channel)} (hend);
\draw[red, fill=white, line width=1pt] (hend) circle (2.5pt);

\draw[->, line width=1pt] (5.2,0) --
      node[above, font=\scriptsize, align=center] {reinterpret as\\[-1pt]off-shell $H^\dagger$} (7.0,0);

\node[draw, rounded corners, fill=blue!5, align=center,
      minimum width=5.2cm, minimum height=1.0cm] (off) at (10.2,0)
      {$\mathcal O_{\rm off}\sim (D^2 H^\dagger)\,(d_{\mathbb{C}}^\dagger L^\dagger)(L^\dagger Q^\dagger)$};
\node[blue!60, font=\scriptsize] at (10.2,0.78) {off-shell $H^\dagger$ insertion};
\draw[->, line width=1pt] (10.2,-0.75) --
      node[right, font=\footnotesize] {$\mathrm{EOM}_{H^\dagger}$} (10.2,-1.85);
\node[draw, rounded corners, fill=green!5, align=center,
      minimum width=5.2cm, minimum height=1.2cm] (sixf) at (10.2,-2.75)
      {$\mathcal O_{6F}\sim (d_{\mathbb{C}}^\dagger L^\dagger)(L^\dagger Q^\dagger)$\\[1pt]
       $\times\bigl(Y_d^\dagger\,Q^\dagger d_{\mathbb{C}}^\dagger + Y_e^\dagger\,L^\dagger e_{\mathbb{C}}^\dagger\bigr)$};
\node[green!40!black, font=\scriptsize] at (10.2,-3.65) {not counted as additional heavy multiplets};
\end{tikzpicture}}
\caption{Illustrative Higgs-like SM channel in a Topology~II factorization of the operator type $d_{\mathbb{C}}^{\dagger 2}L^{\dagger 2}Q^{\dagger 2}$.
The bilinears $L^\dagger Q^\dagger$ and $d_{\mathbb{C}}^\dagger L^\dagger$ are matched to $S_{15}^\dagger(\bar{\mathbf{3}},\mathbf{1},1/3)$ and $S_{20}(\mathbf{3},\mathbf{2},1/6)$, while the color-singlet part of $d_{\mathbb{C}}^\dagger Q^\dagger$ is carried by the Higgs-like channel $S_4(\mathbf{1},\mathbf{2},1/2)\sim H$.
Reinterpreting this line as an off-shell $H^\dagger$ insertion and applying the Higgs equation of motion returns the short-range branch shown in the green box, whose $Y_d^\dagger$ term reproduces $d_{\mathbb{C}}^{\dagger 2}L^{\dagger 2}Q^{\dagger 2}$.  
}
\label{fig:dim9_eom_example}
\end{figure}

The aforementioned example illustrates the reduction of an SM-like insertion, although not every SM-like channel is of this kind.
In reducing to the minimal UV set, having the gauge quantum numbers of an SM multiplet is a necessary, but not sufficient, criterion for removal.
An SM-like heavy mediator must nevertheless be retained in the UV theory if replacing it by its SM counterpart in the diagram fails to generate a local amplitude corresponding to the six-fermion operators, or if the original heavy mediator is a fermion and a mass insertion is required in order for the resulting local amplitude to remain at dimension nine.

\paragraph*{Relation to long-range contributions.}
The difference between the present short-range minimal-UV convention and the broken-phase organization of Refs.~\cite{Bonnet:2012kh,Helo:2015fba} is partly tied to the role of Standard-Model-like propagating fields.
In the present work we focus exclusively on the genuine short-range dim-9 six-fermion sector, and we define the minimal ultraviolet content as the new heavy-field content that remains after SM-like fields have been removed by field redefinitions.
By contrast, in the diagrammatic organization of Ref.~\cite{Bonnet:2012kh}, internal lines whose quantum numbers coincide with those of SM fields are not automatically quotiented out, because in certain topologies such branches can also be reinterpreted as contributions to long-range $\onubb$ amplitudes.
Indeed, in their Fig.~2, diagrams~(a)--(c) correspond to long-range contributions with light-neutrino exchange between two point-like vertices, while diagram~(d) represents purely heavy-particle short-range exchange.
From the perspective adopted here, such SM-like contributions do not contribute to the genuinely new heavy-field content of the short-range dim-9 classification, and are therefore removed in the definition of minimal UV realizations.
The resulting difference is thus of organizational convention, rather than a contradiction in the underlying short-range channel analysis.

With this convention in place, we now summarize the heavy-resonance content of the dim-9 operator space before turning to the minimal-UV classification.

\paragraph*{Heavy-resonance content and comparison }
Table~\ref{tab:resosum} lists all heavy mediator species that appear in at least one tree-level completion of the dim-9 $\onubb$ operators.
The list covers the full dim-9 operator space, including sectors beyond the purely short-range six-fermion one; the restriction to the short-range subset is imposed in the subsequent discussion.

\begingroup
\small
\setlength{\LTcapwidth}{\linewidth}
\begin{longtable}{c|c}
\hline
Scalar & \makecell[tl]{
$S_{2}\ (\mathbf{1},\ \mathbf{1},\ 1)$,\ $S_{3}\ (\mathbf{1},\ \mathbf{1},\ 2)$,\ \textcolor{gray}{$S_{4}\ (\mathbf{1},\ \mathbf{2},\ 1/2)$},\ $S_{8}\ (\mathbf{1},\ \mathbf{3},\ 1)$\\
$S_{14}\ (\mathbf{3},\ \mathbf{1},\ -4/3)$,\ $S_{15}\ (\mathbf{3},\ \mathbf{1},\ -1/3)$,\ $S_{16}\ (\mathbf{3},\ \mathbf{1},\ 2/3)$,\ $S_{20}\ (\mathbf{3},\ \mathbf{2},\ 1/6)$\\
$S_{21}\ (\mathbf{3},\ \mathbf{2},\ 7/6)$,\ $S_{22}\ (\mathbf{3},\ \mathbf{3},\ -1/3)$,\ $S_{27}\ (\mathbf{6},\ \mathbf{1},\ -2/3)$,\ $S_{28}\ (\mathbf{6},\ \mathbf{1},\ 1/3)$\\
$S_{29}\ (\mathbf{6},\ \mathbf{1},\ 4/3)$,\ $S_{34}\ (\mathbf{6},\ \mathbf{3},\ 1/3)$,\ $S_{39}\ (\mathbf{8},\ \mathbf{2},\ 1/2)$
}\\
\hline
Fermion & \makecell[tl]{
$F_{1}\ (\mathbf{1},\ \mathbf{1},\ 0)$,\ $F_{3}\ (\mathbf{1},\ \mathbf{2},\ 1/2)$,\ $F_{5}\ (\mathbf{1},\ \mathbf{3},\ 0)$,\ $F_{10}\ (\mathbf{3},\ \mathbf{1},\ -1/3)$\\
$F_{11}\ (\mathbf{3},\ \mathbf{1},\ 2/3)$,\ $F_{12}\ (\mathbf{3},\ \mathbf{2},\ -5/6)$,\ $F_{13}\ (\mathbf{3},\ \mathbf{2},\ 1/6)$,\ $F_{14}\ (\mathbf{3},\ \mathbf{2},\ 7/6)$\\
$F_{16}\ (\mathbf{3},\ \mathbf{3},\ -1/3)$,\ $F_{17}\ (\mathbf{3},\ \mathbf{3},\ 2/3)$,\ $F_{23}\ (\mathbf{6},\ \mathbf{1},\ 1/3)$,\ $F_{24}\ (\mathbf{6},\ \mathbf{2},\ -1/6)$\\
\textcolor{blue}{$F^{(9)}_{3}\ (\mathbf{3},\ \mathbf{1},\ -4/3)$},\ \textcolor{blue}{$F^{(9)}_{4}\ (\mathbf{3},\ \mathbf{1},\ 5/3)$},\ \textcolor{blue}{$F^{(9)}_{5}\ (\mathbf{8},\ \mathbf{1},\ 0)$},\ \textcolor{blue}{$F^{(9)}_{6}\ (\mathbf{8},\ \mathbf{2},\ 1/2)$}\\
\textcolor{blue}{$F^{(9)}_{7}\ (\mathbf{8},\ \mathbf{3},\ 0)$},\ \textcolor{blue}{$F^{(9)}_{8}\ (\mathbf{6},\ \mathbf{2},\ 5/6)$},\ \textcolor{blue}{$F^{(9)}_{9}\ (\mathbf{6},\ \mathbf{3},\ 1/3)$}
}\\
\hline
Vector & \makecell[tl]{
\textcolor{gray}{$V_{1}\ (\mathbf{1},\ \mathbf{1},\ 0)$},\ $V_{2}\ (\mathbf{1},\ \mathbf{1},\ 1)$,\ $V_{5}\ (\mathbf{1},\ \mathbf{2},\ 3/2)$,\ \textcolor{gray}{$V_{6}\ (\mathbf{1},\ \mathbf{3},\ 0)$}\\
$V_{13}\ (\mathbf{3},\ \mathbf{1},\ 2/3)$,\ $V_{16}\ (\mathbf{3},\ \mathbf{2},\ -5/6)$,\ $V_{17}\ (\mathbf{3},\ \mathbf{2},\ 1/6)$,\ $V_{20}\ (\mathbf{3},\ \mathbf{3},\ 2/3)$\\
$V_{27}\ (\mathbf{6},\ \mathbf{2},\ -1/6)$,\ $V_{28}\ (\mathbf{6},\ \mathbf{2},\ 5/6)$,\ \textcolor{gray}{$V_{30}\ (\mathbf{8},\ \mathbf{1},\ 0)$},\ $V_{31}\ (\mathbf{8},\ \mathbf{1},\ 1)$\\
$V_{32}\ (\mathbf{8},\ \mathbf{3},\ 0)$
}\\
\hline
\caption{Heavy resonances in the pure short-range dim-9 six-fermion sector. Blue entries first appear at dim~9, while gray bosonic entries carry Standard-Model quantum numbers. Parentheses give the $SU(3)_C$ and $SU(2)_L$ representation dimensions together with the $U(1)_Y$ hypercharge.}
\label{tab:resosum}
\end{longtable}
\endgroup

The resonances shown in blue were absent in the previous analysis of effective operators at dimension eight and below.
Their appearance only at dimension nine is itself phenomenologically notable, since it is only at this order that these mediator quantum numbers first enter the $\onubb$ problem.
Several of the genuinely new fermionic entries carry nontrivial color representations, including the octet states $F^{(9)}_{5}(\mathbf{8},\mathbf{1},0)$, $F^{(9)}_{6}(\mathbf{8},\mathbf{2},1/2)$, and $F^{(9)}_{7}(\mathbf{8},\mathbf{3},0)$, as well as the sextet states $F^{(9)}_{8}(\mathbf{6},\mathbf{2},5/6)$ and $F^{(9)}_{9}(\mathbf{6},\mathbf{3},1/3)$.
A particularly useful example is the color-octet fermion $F^{(9)}_{5}(\mathbf{8},\mathbf{1},0)$, whose Standard-Model quantum numbers coincide with those of the gluino.
In the vector sector, some of these channels point rather directly to familiar ultraviolet frameworks, such as the extra gauge bosons of left-right symmetric models or vector leptoquark models; see, for example, Refs.~\cite{Li:2020LRSM0nubb,Li:2023VLQ0nubb}.
These examples show that classifying the quantum numbers of heavy particles before electroweak symmetry breaking helps identify their embedding in well-known UV-complete models.

The detailed resonance-to-operator maps for this pure short-range sector are collected in Appendix~\ref{app:short_resonances}.
As a controlled cross-check, we briefly record the comparison with the broken-phase classification of Ref.~\cite{Bonnet:2012kh} in the same Topology-I notation.
After the strict first-generation projection relevant to the physical $uu\bar{d}\bar{d}\,ee$ channel, the genuine short-range scalar sector is fully reproduced in both the SSS and SFS channels.
The additional content of the present analysis is the electroweak-level treatment of the surviving massive-vector sector and the organization of the short-range catalogue in terms of \emph{minimal UV realizations}, whose precise definition is given in the next subsection.

In addition, we provide a \texttt{Mathematica} function, \texttt{GetUVsForType}, as an add-on to the basis-construction program \texttt{ABC4EFT}, which is available in the \href{https://github.com/haolinli1991/GetUVsForType/tree/main}{GitHub repository}. The function incorporates all the steps shown in Figure~\ref{fig:jbasis_flowchart}, as well as the subtleties related to field redefinitions discussed in this section, and iterates over all possible partitions to generate UV models. The output is given in the form of a list of new fields together with a list of interaction vertices.
Moreover, the function is not restricted to dimension-nine SMEFT operators; in principle, it can be applied to general SMEFT operators containing up to eight fields. Details of its usage, including the input and output formats, are provided in the example notebook in the repository.

\subsection{Minimal ultraviolet realizations}
\label{sec:minimum-uv}

Constructing a UV completion requires specifying the quantum numbers and renormalizable interactions of new fields. Since the coupling structures are typically fixed by gauge symmetry and spin, the minimality of a UV model is governed by the number of unique particle species required, rather than the number of propagators in a particular Feynman topology.
The concept of a minimal UV realization arises because the initial list of completions may overcount the genuinely new heavy content in several ways.
\begin{itemize}
    \item First, a single UV degree of freedom may appear multiple times within a given diagram, manifesting as either the field itself or its conjugate. For example, a configuration derived from a J-basis might involve mediators such as ${A,B,B}$ or ${A,B,B^\dagger}$; we consolidate these into the set $\{A,B\}$, as the primary information for new physics searches is the required set of distinct heavy multiplets.
    \item Second, as discussed previously, SM-like channels do not strictly necessitate genuine heavy degrees of freedom. Such mediators are redundant in our framework and are thus removed from the minimal UV realizations.
    \item Third, we apply a minimality filter based on subset relations. If a UV realization $\{A,B\}$ is already sufficient to generate a given operator, a completion involving an additional degree of freedom, such as $\{A,B,C\}$, is considered non-minimal and is excluded from our list of minimal UV realizations.
\end{itemize}

To illustrate this classification concretely, Table~\ref{tab:short_minimum_uv} displays a few representative completion families. In each block, the first row identifies a minimal UV core, while the subsequent gray rows list non-minimal supersets constructed by appending additional mediators to that same heavy-field set.
This organization highlights the nested structure of the UV realizations, where more complex configurations are shown to be redundant extensions of the underlying minimal core.

\begin{table}[tbp]
\centering
\footnotesize
\setlength{\tabcolsep}{4pt}
\renewcommand{\arraystretch}{1.08}
\begin{tabular*}{\textwidth}{@{\extracolsep{\fill}}>{\raggedright\arraybackslash}p{0.27\textwidth}>{%
\raggedright\arraybackslash}p{0.22\textwidth}>{%
\raggedright\arraybackslash}p{0.22\textwidth}>{%
\raggedright\arraybackslash}p{0.22\textwidth}@{}}
\toprule
\textbf{Operator} & \textbf{Mediator 1} & \textbf{Mediator 2} & \textbf{Mediator 3} \\
\midrule
\(d_{\mathbb C}^{\dagger} L^{\dagger 2} Q Q^{\dagger} u_{\mathbb C}\) & $F_{10}\ (\mathbf{3},\ \mathbf{1},\ -1/3)$ & $V_{17}\ (\mathbf{3},\ \mathbf{2},\ 1/6)$ &  \\
 & \textcolor{gray}{$F_{10}\ (\mathbf{3},\ \mathbf{1},\ -1/3)$} & \textcolor{gray}{$V_{13}\ (\mathbf{3},\ \mathbf{1},\ 2/3)$} & \textcolor{gray}{$V_{17}\ (\mathbf{3},\ \mathbf{2},\ 1/6)$} \\
 & \textcolor{gray}{$F_{10}\ (\mathbf{3},\ \mathbf{1},\ -1/3)$} & \textcolor{gray}{$V_{16}\ (\mathbf{3},\ \mathbf{2},\ -5/6)$} & \textcolor{gray}{$V_{17}\ (\mathbf{3},\ \mathbf{2},\ 1/6)$} \\
 & \textcolor{gray}{$S_{15}\ (\mathbf{3},\ \mathbf{1},\ -1/3)$} & \textcolor{gray}{$F_{10}\ (\mathbf{3},\ \mathbf{1},\ -1/3)$} & \textcolor{gray}{$V_{17}\ (\mathbf{3},\ \mathbf{2},\ 1/6)$} \\
 & \textcolor{gray}{$S_{20}\ (\mathbf{3},\ \mathbf{2},\ 1/6)$} & \textcolor{gray}{$F_{10}\ (\mathbf{3},\ \mathbf{1},\ -1/3)$} & \textcolor{gray}{$V_{17}\ (\mathbf{3},\ \mathbf{2},\ 1/6)$} \\
\midrule
\(d_{\mathbb C}^{\dagger} L^{\dagger 2} Q Q^{\dagger} u_{\mathbb C}\)
 & $S_{4}\ (\mathbf{1},\ \mathbf{2},\ 1/2)$ & $F_{5}\ (\mathbf{1},\ \mathbf{3},\ 0)$ &  \\
 & \textcolor{gray}{$S_{4}\ (\mathbf{1},\ \mathbf{2},\ 1/2)$} & \textcolor{gray}{$F_{5}\ (\mathbf{1},\ \mathbf{3},\ 0)$} & \textcolor{gray}{$V_{17}\ (\mathbf{3},\ \mathbf{2},\ 1/6)$} \\
 & \textcolor{gray}{$S_{4}\ (\mathbf{1},\ \mathbf{2},\ 1/2)$} & \textcolor{gray}{$F_{5}\ (\mathbf{1},\ \mathbf{3},\ 0)$} & \textcolor{gray}{$V_{20}\ (\mathbf{3},\ \mathbf{3},\ 2/3)$} \\
 & \textcolor{gray}{$S_{4}\ (\mathbf{1},\ \mathbf{2},\ 1/2)$} & \textcolor{gray}{$S_{20}\ (\mathbf{3},\ \mathbf{2},\ 1/6)$} & \textcolor{gray}{$F_{5}\ (\mathbf{1},\ \mathbf{3},\ 0)$} \\
 & \textcolor{gray}{$S_{4}\ (\mathbf{1},\ \mathbf{2},\ 1/2)$} & \textcolor{gray}{$S_{22}\ (\mathbf{3},\ \mathbf{3},\ -1/3)$} & \textcolor{gray}{$F_{5}\ (\mathbf{1},\ \mathbf{3},\ 0)$} \\
\midrule
\(d_{\mathbb C}^{\dagger} e_{\mathbb C} L^{\dagger} Q u_{\mathbb C}^{2}\) & $S_{29}\ (\mathbf{6},\ \mathbf{1},\ 4/3)$ & $V_{13}\ (\mathbf{3},\ \mathbf{1},\ 2/3)$ &  \\
 & \textcolor{gray}{$S_{29}\ (\mathbf{6},\ \mathbf{1},\ 4/3)$} & \textcolor{gray}{$F^{(9)}_{4}\ (\mathbf{3},\ \mathbf{1},\ 5/3)$} & \textcolor{gray}{$V_{13}\ (\mathbf{3},\ \mathbf{1},\ 2/3)$} \\
 & \textcolor{gray}{$S_{29}\ (\mathbf{6},\ \mathbf{1},\ 4/3)$} & \textcolor{gray}{$F^{(9)}_{8}\ (\mathbf{6},\ \mathbf{2},\ 5/6)$} & \textcolor{gray}{$V_{13}\ (\mathbf{3},\ \mathbf{1},\ 2/3)$} \\
 & \textcolor{gray}{$S_{29}\ (\mathbf{6},\ \mathbf{1},\ 4/3)$} & \textcolor{gray}{$F_{14}\ (\mathbf{3},\ \mathbf{2},\ 7/6)$} & \textcolor{gray}{$V_{13}\ (\mathbf{3},\ \mathbf{1},\ 2/3)$} \\
 & \textcolor{gray}{$S_{29}\ (\mathbf{6},\ \mathbf{1},\ 4/3)$} & \textcolor{gray}{$F_{23}\ (\mathbf{6},\ \mathbf{1},\ 1/3)$} & \textcolor{gray}{$V_{13}\ (\mathbf{3},\ \mathbf{1},\ 2/3)$} \\
\midrule
\(d_{\mathbb C}^{\dagger 2} e_{\mathbb C} L^{\dagger} Q^{\dagger} u_{\mathbb C}\) & $S_{15}\ (\mathbf{3},\ \mathbf{1},\ -1/3)$ & $S_{27}\ (\mathbf{6},\ \mathbf{1},\ -2/3)$ &  \\
 & \textcolor{gray}{$S_{15}\ (\mathbf{3},\ \mathbf{1},\ -1/3)$} & \textcolor{gray}{$S_{27}\ (\mathbf{6},\ \mathbf{1},\ -2/3)$} & \textcolor{gray}{$F^{(9)}_{3}\ (\mathbf{3},\ \mathbf{1},\ -4/3)$} \\
 & \textcolor{gray}{$S_{15}\ (\mathbf{3},\ \mathbf{1},\ -1/3)$} & \textcolor{gray}{$S_{27}\ (\mathbf{6},\ \mathbf{1},\ -2/3)$} & \textcolor{gray}{$F_{12}\ (\mathbf{3},\ \mathbf{2},\ -5/6)$} \\
 & \textcolor{gray}{$S_{15}\ (\mathbf{3},\ \mathbf{1},\ -1/3)$} & \textcolor{gray}{$S_{27}\ (\mathbf{6},\ \mathbf{1},\ -2/3)$} & \textcolor{gray}{$F_{23}\ (\mathbf{6},\ \mathbf{1},\ 1/3)$} \\
 & \textcolor{gray}{$S_{15}\ (\mathbf{3},\ \mathbf{1},\ -1/3)$} & \textcolor{gray}{$S_{27}\ (\mathbf{6},\ \mathbf{1},\ -2/3)$} & \textcolor{gray}{$F_{24}\ (\mathbf{6},\ \mathbf{2},\ -1/6)$} \\
\bottomrule
\end{tabular*}
\caption{Sample UV completion families from the strict first-generation short-range dim-9 classification.
In each block, the first row is a minimal-UV classification and the gray rows list non-minimal supersets built on the same family.}
\label{tab:short_minimum_uv}
\end{table}

Table~\ref{tab:minuv_summary} provides a systematic count of the UV realizations for the six short-range dimension-nine operator types. We characterize each realization by $N_{\rm med}$, the number of distinct heavy multiplets required after applying the minimality criteria discussed above. A realization is classified as minimal if no proper subset of its heavy-field content is sufficient to generate the same short-range operator type.

Applying these criteria to the first-generation SMEFT operators, we identify 440 minimal UV realizations out of a total of 505 unique mediator combinations. The reduction from 505 to 440 reflects the elimination of non-minimal cases—specifically, 77 realizations that effectively reduce to a more economical set of only 12 distinct two-mediator combinations ($N_{\rm med}=2$). The vast majority of the minimal realizations (428 out of 440) strictly necessitate three distinct heavy species ($N_{\rm med}=3$). Notably, no $N_{\rm med}=1$ case survives the requirement of a genuine six-fermion tree-level completion.

The operator-by-operator breakdown reveals that most classes are driven toward $N_{\rm med}=3$ once the full first-generation flavor, color, and $SU(2)_L$ constraints are imposed. Only certain mixed-chirality operators retain a small number of $N_{\rm med}=2$ realizations, whereas other classes exclusively require three distinct mediators. In this framework, $N_{\rm med}$ serves as a direct measure of the UV economy for each operator type, indicating the minimum beyond-SM complexity required for its generation.

\begin{table}[tbp]
\centering
\small
\setlength{\tabcolsep}{4.5pt}
\renewcommand{\arraystretch}{1.08}
\begin{tabularx}{\textwidth}{@{}>{\raggedright\arraybackslash}Xccccc@{}}
\toprule
\textbf{Operator type} & \textbf{UV classifications} & \textbf{Minimal UV} & \textbf{$N_{\rm med}=2$} & \textbf{$N_{\rm med}=3$} & \textbf{Min.\ UV with $V$} \\
\midrule
$L^{\dagger 2}Q^{2}u_{\mathbb C}^{2}$ & 40 & 40 & 0 & 40 & 32 \\
$d_{\mathbb C}^\dagger L^{\dagger 2}QQ^\dagger u_{\mathbb C}$ & 195 & 153 & 8 & 145 & 127 \\
$d_{\mathbb C}^\dagger e_{\mathbb C}L^\dagger Q u_{\mathbb C}^{2}$ & 102 & 83 & 3 & 80 & 62 \\
$d_{\mathbb C}^{\dagger 2}L^{\dagger 2}Q^{\dagger 2}$ & 40 & 40 & 0 & 40 & 0 \\
$d_{\mathbb C}^{\dagger 2}e_{\mathbb C}L^\dagger Q^\dagger u_{\mathbb C}$ & 102 & 98 & 1 & 97 & 83 \\
$d_{\mathbb C}^{\dagger 2}e_{\mathbb C}^{2}u_{\mathbb C}^{2}$ & 26 & 26 & 0 & 26 & 20 \\
\midrule
\textbf{Total} & \textbf{505} & \textbf{440} & \textbf{12} & \textbf{428} & \textbf{324} \\
\bottomrule
\end{tabularx}
\caption{Summary of the minimal-UV counting for the short-range dim-9 classification. }
\label{tab:minuv_summary}
\end{table}

The complete operator-by-operator classification, organized by minimal UV cores and their non-minimal extensions, is collected in Appendix~\ref{app:full_short_catalogue}.
In this form, the catalog serves as a practical starting point for ultraviolet model building: one may first choose the minimal UV core compatible with a target operator type, and subsequently incorporate non-minimal extensions as required by specific model assumptions.
This result thus provides a systematic reference for connecting short-range $\onubb$ operators to their potential BSM origins.

\section{Summary and Conclusions }
\label{sec:concl}

In this work, we have performed a systematic classification of tree-level UV completions for short-range $0\nu\beta\beta$ based on $d=9$ six-fermion effective operators within the SMEFT framework. 
With the J-basis analysis, we established a rigorous mapping between the independent operator basis and their tree-level UV origins, indexed by the spin and gauge representations of the channels that define each tree topology. 
By working in the unbroken phase, our analysis identifies the genuine heavy degrees of freedom most relevant to future collider searches.
To ensure phenomenological consistency with low-energy $0\nu\beta\beta$ processes, we focus on first-generation SMEFT operators that match onto the $uu\bar{d}\bar{d}ee$ sector of the LEFT.

We presented a comprehensive compilation of 505 distinct combinations of mediators. 
We merge diverse tree-level topologies into single instances whenever they utilize the same set of heavy fields, as the specific internal connection patterns are of secondary importance for a model-independent survey.
Requiring a minimal UV realization, in the sense that removing any mediator from the list prevents the generation of the operator of interest, we find that 428 cases are irreducible, while the remaining 77 reduce to 12 distinct two-mediator combinations.
Notably, while our results reproduce the all-scalar and scalar-fermion-scalar completions found in Ref.~\cite{Bonnet:2012kh} and complement the vector analysis of Ref.~\cite{Fonseca:2016tbn}, this study identifies 324 minimal UV configurations featuring vector resonances within a complete short-range dim-9 SMEFT classification. 
This majority share suggests that new physics models incorporating heavy vectors represent a significant class of theories capable of generating short-range $0\nu\beta\beta$.

\section*{Acknowledgements}
H.-L.L. is supported by the National Natural Science Foundation of China under Grant No. 1250050417 and by the start-up funding of Sun Yat-Sen University under Grant No. 74130-12255013.
M.-L.X. is supported by the National Natural Science Foundation of China under Grant No. 12405123, by the Fundamental Research Funds for the Central Universities at Sun Yat-Sen University under Grant No. 25hytd001, and by the Shenzhen Science and Technology Program under Grant No. JCYJ20240813150911015.
Y.-H.N. is supported by the University Development Fund of The Chinese University of Hong Kong, Shenzhen, under Grant No. UDF01003912.
J.-H.Y. is supported by the National Natural Science Foundation of China under Grants No. 12347105, 12375099, and 12447101, and by the National Key Research and Development Program of China under Grants No. 2020YFC2201501 and 2021YFA0718304.
\appendix
\section{Detailed short-range resonance tables}
\label{app:short_resonances}
For completeness we collect here the resonance-to-operator maps for the pure short-range dim-9 sector, namely the operator types containing six fermion fields only. The operator shorthand is the same as in Sec.~\ref{sec:eft}, and the resonance labels are aligned with the summary list in Table~\ref{tab:resosum}.
\setlength{\LTcapwidth}{\linewidth}

\begin{longtable}{c|p{0.65\textwidth}}
    \toprule
    \multicolumn{2}{c}{Scalar}\\
    \midrule
    \multicolumn{1}{c}{$\left(SU(3)_C,\ SU(2)_L,\ U(1)_Y\right)$} & \multicolumn{1}{c}{Operator types} \\
    \midrule
    \endhead
  $S_{2}\ (\mathbf{1},\ \mathbf{1},\ 1)$ & $L^{\dagger 2} Q^{2} u_{\mathbb{C}}^{2}$\ \ $d_{\mathbb{C}}^\dagger L^{\dagger 2} Q Q^\dagger u_{\mathbb{C}}$\ \ $d_{\mathbb{C}}^{\dagger 2} L^{\dagger 2} Q^{\dagger 2}$\\
  \midrule
  $S_{3}\ (\mathbf{1},\ \mathbf{1},\ 2)$ & $d_{\mathbb{C}}^{\dagger 2} e_{\mathbb{C}}^{2} u_{\mathbb{C}}^{2}$\\
  \midrule
  $S_{4}\ (\mathbf{1},\ \mathbf{2},\ 1/2)$ & $L^{\dagger 2} Q^{2} u_{\mathbb{C}}^{2}$\ \ $d_{\mathbb{C}}^\dagger L^{\dagger 2} Q Q^\dagger u_{\mathbb{C}}$\ \ $d_{\mathbb{C}}^\dagger e_{\mathbb{C}} L^\dagger Q u_{\mathbb{C}}^{2}$\ \ $d_{\mathbb{C}}^{\dagger 2} L^{\dagger 2} Q^{\dagger 2}$\ \ $d_{\mathbb{C}}^{\dagger 2} e_{\mathbb{C}} L^\dagger Q^\dagger u_{\mathbb{C}}$\\
  \midrule
  $S_{8}\ (\mathbf{1},\ \mathbf{3},\ 1)$ & $L^{\dagger 2} Q^{2} u_{\mathbb{C}}^{2}$\ \ $d_{\mathbb{C}}^\dagger L^{\dagger 2} Q Q^\dagger u_{\mathbb{C}}$\ \ $d_{\mathbb{C}}^{\dagger 2} L^{\dagger 2} Q^{\dagger 2}$\\
  \midrule
  $S_{14}\ (\mathbf{3},\ \mathbf{1},\ -4/3)$ & $L^{\dagger 2} Q^{2} u_{\mathbb{C}}^{2}$\ \ $d_{\mathbb{C}}^\dagger e_{\mathbb{C}} L^\dagger Q u_{\mathbb{C}}^{2}$\ \ $d_{\mathbb{C}}^{\dagger 2} e_{\mathbb{C}}^{2} u_{\mathbb{C}}^{2}$\\
  \midrule
  $S_{15}\ (\mathbf{3},\ \mathbf{1},\ -1/3)$ & $L^{\dagger 2} Q^{2} u_{\mathbb{C}}^{2}$\ \ $d_{\mathbb{C}}^\dagger L^{\dagger 2} Q Q^\dagger u_{\mathbb{C}}$\ \ $d_{\mathbb{C}}^\dagger e_{\mathbb{C}} L^\dagger Q u_{\mathbb{C}}^{2}$\ \ $d_{\mathbb{C}}^{\dagger 2} L^{\dagger 2} Q^{\dagger 2}$\ \ $d_{\mathbb{C}}^{\dagger 2} e_{\mathbb{C}} L^\dagger Q^\dagger u_{\mathbb{C}}$\ \ $d_{\mathbb{C}}^{\dagger 2} e_{\mathbb{C}}^{2} u_{\mathbb{C}}^{2}$\\
  \midrule
  $S_{16}\ (\mathbf{3},\ \mathbf{1},\ 2/3)$ & $d_{\mathbb{C}}^{\dagger 2} L^{\dagger 2} Q^{\dagger 2}$\ \ $d_{\mathbb{C}}^{\dagger 2} e_{\mathbb{C}} L^\dagger Q^\dagger u_{\mathbb{C}}$\ \ $d_{\mathbb{C}}^{\dagger 2} e_{\mathbb{C}}^{2} u_{\mathbb{C}}^{2}$\\
\end{longtable}

\addtocounter{table}{-1}
\begin{longtable}{c|p{0.65\textwidth}}
    \toprule
    \multicolumn{2}{c}{Fermion}\\
    \midrule
    \multicolumn{1}{c}{$\left(SU(3)_C,\ SU(2)_L,\ U(1)_Y\right)$} & \multicolumn{1}{c}{Operator types} \\
    \midrule
    \endhead
  $F_{1}\ (\mathbf{1},\ \mathbf{1},\ 0)$ & $L^{\dagger 2} Q^{2} u_{\mathbb{C}}^{2}$\ \ $d_{\mathbb{C}}^\dagger L^{\dagger 2} Q Q^\dagger u_{\mathbb{C}}$\ \ $d_{\mathbb{C}}^\dagger e_{\mathbb{C}} L^\dagger Q u_{\mathbb{C}}^{2}$\ \ $d_{\mathbb{C}}^{\dagger 2} L^{\dagger 2} Q^{\dagger 2}$\ \ $d_{\mathbb{C}}^{\dagger 2} e_{\mathbb{C}} L^\dagger Q^\dagger u_{\mathbb{C}}$\ \ $d_{\mathbb{C}}^{\dagger 2} e_{\mathbb{C}}^{2} u_{\mathbb{C}}^{2}$\\
  \midrule
  $F_{3}\ (\mathbf{1},\ \mathbf{2},\ 1/2)$ & $d_{\mathbb{C}}^\dagger L^{\dagger 2} Q Q^\dagger u_{\mathbb{C}}$\ \ $d_{\mathbb{C}}^\dagger e_{\mathbb{C}} L^\dagger Q u_{\mathbb{C}}^{2}$\ \ $d_{\mathbb{C}}^{\dagger 2} e_{\mathbb{C}} L^\dagger Q^\dagger u_{\mathbb{C}}$\\
  \midrule
  $F_{5}\ (\mathbf{1},\ \mathbf{3},\ 0)$ & $L^{\dagger 2} Q^{2} u_{\mathbb{C}}^{2}$\ \ $d_{\mathbb{C}}^\dagger L^{\dagger 2} Q Q^\dagger u_{\mathbb{C}}$\ \ $d_{\mathbb{C}}^{\dagger 2} L^{\dagger 2} Q^{\dagger 2}$\\
  \midrule
  $F^{(9)}_{3}\ (\mathbf{3},\ \mathbf{1},\ -4/3)$ & $d_{\mathbb{C}}^{\dagger 2} e_{\mathbb{C}} L^\dagger Q^\dagger u_{\mathbb{C}}$\ \ $d_{\mathbb{C}}^{\dagger 2} e_{\mathbb{C}}^{2} u_{\mathbb{C}}^{2}$\\
  \midrule
  $F_{10}\ (\mathbf{3},\ \mathbf{1},\ -1/3)$ & $L^{\dagger 2} Q^{2} u_{\mathbb{C}}^{2}$\ \ $d_{\mathbb{C}}^\dagger L^{\dagger 2} Q Q^\dagger u_{\mathbb{C}}$\ \ $d_{\mathbb{C}}^\dagger e_{\mathbb{C}} L^\dagger Q u_{\mathbb{C}}^{2}$\ \ $d_{\mathbb{C}}^{\dagger 2} e_{\mathbb{C}} L^\dagger Q^\dagger u_{\mathbb{C}}$\ \ $d_{\mathbb{C}}^{\dagger 2} e_{\mathbb{C}}^{2} u_{\mathbb{C}}^{2}$\\
  \midrule
  $F_{11}\ (\mathbf{3},\ \mathbf{1},\ 2/3)$ & $d_{\mathbb{C}}^\dagger L^{\dagger 2} Q Q^\dagger u_{\mathbb{C}}$\ \ $d_{\mathbb{C}}^{\dagger 2} L^{\dagger 2} Q^{\dagger 2}$\\
  \midrule
  $F^{(9)}_{4}\ (\mathbf{3},\ \mathbf{1},\ 5/3)$ & $d_{\mathbb{C}}^\dagger e_{\mathbb{C}} L^\dagger Q u_{\mathbb{C}}^{2}$\ \ $d_{\mathbb{C}}^{\dagger 2} e_{\mathbb{C}}^{2} u_{\mathbb{C}}^{2}$\\
\end{longtable}

\addtocounter{table}{-1}
\begin{longtable}{c|p{0.65\textwidth}}
    \toprule
    \multicolumn{2}{c}{Vector}\\
    \midrule
    \multicolumn{1}{c}{$\left(SU(3)_C,\ SU(2)_L,\ U(1)_Y\right)$} & \multicolumn{1}{c}{Operator types} \\
    \midrule
    \endhead
  $V_{1}\ (\mathbf{1},\ \mathbf{1},\ 0)$ & $d_{\mathbb{C}}^\dagger L^{\dagger 2} Q Q^\dagger u_{\mathbb{C}}$\\
  \midrule
  $V_{2}\ (\mathbf{1},\ \mathbf{1},\ 1)$ & $d_{\mathbb{C}}^\dagger L^{\dagger 2} Q Q^\dagger u_{\mathbb{C}}$\ \ $d_{\mathbb{C}}^\dagger e_{\mathbb{C}} L^\dagger Q u_{\mathbb{C}}^{2}$\ \ $d_{\mathbb{C}}^{\dagger 2} e_{\mathbb{C}} L^\dagger Q^\dagger u_{\mathbb{C}}$\ \ $d_{\mathbb{C}}^{\dagger 2} e_{\mathbb{C}}^{2} u_{\mathbb{C}}^{2}$\\
  \midrule
  $V_{5}\ (\mathbf{1},\ \mathbf{2},\ 3/2)$ & $d_{\mathbb{C}}^\dagger e_{\mathbb{C}} L^\dagger Q u_{\mathbb{C}}^{2}$\ \ $d_{\mathbb{C}}^{\dagger 2} e_{\mathbb{C}} L^\dagger Q^\dagger u_{\mathbb{C}}$\\
  \midrule
  $V_{6}\ (\mathbf{1},\ \mathbf{3},\ 0)$ & $d_{\mathbb{C}}^\dagger L^{\dagger 2} Q Q^\dagger u_{\mathbb{C}}$\\
  \midrule
  $V_{13}\ (\mathbf{3},\ \mathbf{1},\ 2/3)$ & $L^{\dagger 2} Q^{2} u_{\mathbb{C}}^{2}$\ \ $d_{\mathbb{C}}^\dagger L^{\dagger 2} Q Q^\dagger u_{\mathbb{C}}$\ \ $d_{\mathbb{C}}^\dagger e_{\mathbb{C}} L^\dagger Q u_{\mathbb{C}}^{2}$\ \ $d_{\mathbb{C}}^{\dagger 2} e_{\mathbb{C}} L^\dagger Q^\dagger u_{\mathbb{C}}$\ \ $d_{\mathbb{C}}^{\dagger 2} e_{\mathbb{C}}^{2} u_{\mathbb{C}}^{2}$\\
  \midrule
  $V_{16}\ (\mathbf{3},\ \mathbf{2},\ -5/6)$ & $d_{\mathbb{C}}^\dagger L^{\dagger 2} Q Q^\dagger u_{\mathbb{C}}$\ \ $d_{\mathbb{C}}^{\dagger 2} e_{\mathbb{C}} L^\dagger Q^\dagger u_{\mathbb{C}}$\\
  \midrule
  $V_{17}\ (\mathbf{3},\ \mathbf{2},\ 1/6)$ & $L^{\dagger 2} Q^{2} u_{\mathbb{C}}^{2}$\ \ $d_{\mathbb{C}}^\dagger L^{\dagger 2} Q Q^\dagger u_{\mathbb{C}}$\ \ $d_{\mathbb{C}}^\dagger e_{\mathbb{C}} L^\dagger Q u_{\mathbb{C}}^{2}$\ \ $d_{\mathbb{C}}^{\dagger 2} e_{\mathbb{C}} L^\dagger Q^\dagger u_{\mathbb{C}}$\\
  \midrule
  $V_{20}\ (\mathbf{3},\ \mathbf{3},\ 2/3)$ & $L^{\dagger 2} Q^{2} u_{\mathbb{C}}^{2}$\ \ $d_{\mathbb{C}}^\dagger L^{\dagger 2} Q Q^\dagger u_{\mathbb{C}}$\\
\end{longtable}

\clearpage
\section{Complete short-range UV catalogue by operator type}
\label{app:full_short_catalogue}
The complete short-range dim-9 catalogue is listed below by SMEFT operator type, with the first row in each block giving a minimal UV realization and the gray rows listing its non-minimal supersets.
\setlength{\LTcapwidth}{\linewidth}

\begingroup
\small
\renewcommand{\arraystretch}{1.08}
\begin{longtable}{@{\hspace{1.2em}}>{\raggedright\arraybackslash}p{0.255\textwidth}@{\hspace{1.15em}}>{\raggedright\arraybackslash}p{0.255\textwidth}@{\hspace{1.15em}}>{\raggedright\arraybackslash}p{0.255\textwidth}@{}}
\hline
\multicolumn{3}{@{}l@{}}{{\normalsize\bfseries Operator type:} $L^{\dagger 2} Q^{2} u_{\mathbb C}^{2}$} \\
\hline
\textbf{Mediator 1} & \textbf{Mediator 2} & \textbf{Mediator 3} \\
\hline
\endfirsthead
\hline
\multicolumn{3}{@{}l@{}}{{\normalsize\bfseries Operator type:} $L^{\dagger 2} Q^{2} u_{\mathbb C}^{2}$} \\
\hline
\textbf{Mediator 1} & \textbf{Mediator 2} & \textbf{Mediator 3} \\
\hline
\endhead
\hline
\endfoot
$F^{(9)}_{5}\ (\mathbf{8},\ \mathbf{1},\ 0)$ & $V_{13}\ (\mathbf{3},\ \mathbf{1},\ 2/3)$ & $V_{17}\ (\mathbf{3},\ \mathbf{2},\ 1/6)$ \\
\cmidrule(lr){1-3}
$F^{(9)}_{7}\ (\mathbf{8},\ \mathbf{3},\ 0)$ & $V_{17}\ (\mathbf{3},\ \mathbf{2},\ 1/6)$ & $V_{20}\ (\mathbf{3},\ \mathbf{3},\ 2/3)$ \\
\cmidrule(lr){1-3}
$F^{(9)}_{8}\ (\mathbf{6},\ \mathbf{2},\ 5/6)$ & $V_{13}\ (\mathbf{3},\ \mathbf{1},\ 2/3)$ & $V_{17}\ (\mathbf{3},\ \mathbf{2},\ 1/6)$ \\
\cmidrule(lr){1-3}
$F^{(9)}_{8}\ (\mathbf{6},\ \mathbf{2},\ 5/6)$ & $V_{17}\ (\mathbf{3},\ \mathbf{2},\ 1/6)$ & $V_{20}\ (\mathbf{3},\ \mathbf{3},\ 2/3)$ \\
\cmidrule(lr){1-3}
$F_{12}\ (\mathbf{3},\ \mathbf{2},\ -5/6)$ & $V_{13}\ (\mathbf{3},\ \mathbf{1},\ 2/3)$ & $V_{17}\ (\mathbf{3},\ \mathbf{2},\ 1/6)$ \\
\cmidrule(lr){1-3}
$F_{12}\ (\mathbf{3},\ \mathbf{2},\ -5/6)$ & $V_{17}\ (\mathbf{3},\ \mathbf{2},\ 1/6)$ & $V_{20}\ (\mathbf{3},\ \mathbf{3},\ 2/3)$ \\
\cmidrule(lr){1-3}
$F_{1}\ (\mathbf{1},\ \mathbf{1},\ 0)$ & $V_{13}\ (\mathbf{3},\ \mathbf{1},\ 2/3)$ & $V_{17}\ (\mathbf{3},\ \mathbf{2},\ 1/6)$ \\
\cmidrule(lr){1-3}
$F_{5}\ (\mathbf{1},\ \mathbf{3},\ 0)$ & $V_{17}\ (\mathbf{3},\ \mathbf{2},\ 1/6)$ & $V_{20}\ (\mathbf{3},\ \mathbf{3},\ 2/3)$ \\
\end{longtable}
\refstepcounter{table}
\begin{center}
\small\textbf{Table \thetable.} Short-range UV catalogue for $L^{\dagger 2} Q^{2} u_{\mathbb C}^{2}$.
\label{tab:app_short_catalogue_1}
\end{center}
\endgroup

\begingroup
\small
\renewcommand{\arraystretch}{1.08}
\begin{longtable}{@{\hspace{1.2em}}>{\raggedright\arraybackslash}p{0.255\textwidth}@{\hspace{1.15em}}>{\raggedright\arraybackslash}p{0.255\textwidth}@{\hspace{1.15em}}>{\raggedright\arraybackslash}p{0.255\textwidth}@{}}
\hline
\multicolumn{3}{@{}l@{}}{{\normalsize\bfseries Operator type:} $d_{\mathbb C}^\dagger L^{\dagger 2} Q Q^\dagger u_{\mathbb C}$} \\
\hline
\textbf{Mediator 1} & \textbf{Mediator 2} & \textbf{Mediator 3} \\
\hline
\endfirsthead
\hline
\multicolumn{3}{@{}l@{}}{{\normalsize\bfseries Operator type:} $d_{\mathbb C}^\dagger L^{\dagger 2} Q Q^\dagger u_{\mathbb C}$} \\
\hline
\textbf{Mediator 1} & \textbf{Mediator 2} & \textbf{Mediator 3} \\
\hline
\endhead
\hline
\endfoot
$F_{10}\ (\mathbf{3},\ \mathbf{1},\ -1/3)$ & $V_{17}\ (\mathbf{3},\ \mathbf{2},\ 1/6)$ &  \\
\textcolor{gray}{$F_{10}\ (\mathbf{3},\ \mathbf{1},\ -1/3)$} & \textcolor{gray}{$V_{13}\ (\mathbf{3},\ \mathbf{1},\ 2/3)$} & \textcolor{gray}{$V_{17}\ (\mathbf{3},\ \mathbf{2},\ 1/6)$} \\
\textcolor{gray}{$F_{10}\ (\mathbf{3},\ \mathbf{1},\ -1/3)$} & \textcolor{gray}{$V_{16}\ (\mathbf{3},\ \mathbf{2},\ -5/6)$} & \textcolor{gray}{$V_{17}\ (\mathbf{3},\ \mathbf{2},\ 1/6)$} \\
\textcolor{gray}{$S_{15}\ (\mathbf{3},\ \mathbf{1},\ -1/3)$} & \textcolor{gray}{$F_{10}\ (\mathbf{3},\ \mathbf{1},\ -1/3)$} & \textcolor{gray}{$V_{17}\ (\mathbf{3},\ \mathbf{2},\ 1/6)$} \\
\textcolor{gray}{$S_{20}\ (\mathbf{3},\ \mathbf{2},\ 1/6)$} & \textcolor{gray}{$F_{10}\ (\mathbf{3},\ \mathbf{1},\ -1/3)$} & \textcolor{gray}{$V_{17}\ (\mathbf{3},\ \mathbf{2},\ 1/6)$} \\
\cmidrule(lr){1-3}
$F_{16}\ (\mathbf{3},\ \mathbf{3},\ -1/3)$ & $V_{17}\ (\mathbf{3},\ \mathbf{2},\ 1/6)$ &  \\
\textcolor{gray}{$F_{16}\ (\mathbf{3},\ \mathbf{3},\ -1/3)$} & \textcolor{gray}{$V_{16}\ (\mathbf{3},\ \mathbf{2},\ -5/6)$} & \textcolor{gray}{$V_{17}\ (\mathbf{3},\ \mathbf{2},\ 1/6)$} \\
\textcolor{gray}{$F_{16}\ (\mathbf{3},\ \mathbf{3},\ -1/3)$} & \textcolor{gray}{$V_{17}\ (\mathbf{3},\ \mathbf{2},\ 1/6)$} & \textcolor{gray}{$V_{20}\ (\mathbf{3},\ \mathbf{3},\ 2/3)$} \\
\end{longtable}
\refstepcounter{table}
\begin{center}
\small\textbf{Table \thetable.} Short-range UV catalogue for $d_{\mathbb C}^\dagger L^{\dagger 2} Q Q^\dagger u_{\mathbb C}$.
\label{tab:app_short_catalogue_2}
\end{center}
\endgroup

\begingroup
\small
\renewcommand{\arraystretch}{1.08}
\begin{longtable}{@{\hspace{1.2em}}>{\raggedright\arraybackslash}p{0.255\textwidth}@{\hspace{1.15em}}>{\raggedright\arraybackslash}p{0.255\textwidth}@{\hspace{1.15em}}>{\raggedright\arraybackslash}p{0.255\textwidth}@{}}
\hline
\multicolumn{3}{@{}l@{}}{{\normalsize\bfseries Operator type:} $d_{\mathbb C}^\dagger e_{\mathbb C} L^\dagger Q u_{\mathbb C}^{2}$} \\
\hline
\textbf{Mediator 1} & \textbf{Mediator 2} & \textbf{Mediator 3} \\
\hline
\endfirsthead
\hline
\multicolumn{3}{@{}l@{}}{{\normalsize\bfseries Operator type:} $d_{\mathbb C}^\dagger e_{\mathbb C} L^\dagger Q u_{\mathbb C}^{2}$} \\
\hline
\textbf{Mediator 1} & \textbf{Mediator 2} & \textbf{Mediator 3} \\
\hline
\endhead
\hline
\endfoot
$F_{12}\ (\mathbf{3},\ \mathbf{2},\ -5/6)$ & $V_{17}\ (\mathbf{3},\ \mathbf{2},\ 1/6)$ &  \\
\textcolor{gray}{$F_{12}\ (\mathbf{3},\ \mathbf{2},\ -5/6)$} & \textcolor{gray}{$V_{13}\ (\mathbf{3},\ \mathbf{1},\ 2/3)$} & \textcolor{gray}{$V_{17}\ (\mathbf{3},\ \mathbf{2},\ 1/6)$} \\
\textcolor{gray}{$F_{12}\ (\mathbf{3},\ \mathbf{2},\ -5/6)$} & \textcolor{gray}{$V_{17}\ (\mathbf{3},\ \mathbf{2},\ 1/6)$} & \textcolor{gray}{$V_{27}\ (\mathbf{6},\ \mathbf{2},\ -1/6)$} \\
\textcolor{gray}{$F_{12}\ (\mathbf{3},\ \mathbf{2},\ -5/6)$} & \textcolor{gray}{$V_{17}\ (\mathbf{3},\ \mathbf{2},\ 1/6)$} & \textcolor{gray}{$V_{31}\ (\mathbf{8},\ \mathbf{1},\ 1)$} \\
\textcolor{gray}{$F_{12}\ (\mathbf{3},\ \mathbf{2},\ -5/6)$} & \textcolor{gray}{$V_{2}\ (\mathbf{1},\ \mathbf{1},\ 1)$} & \textcolor{gray}{$V_{17}\ (\mathbf{3},\ \mathbf{2},\ 1/6)$} \\
\textcolor{gray}{$F_{12}\ (\mathbf{3},\ \mathbf{2},\ -5/6)$} & \textcolor{gray}{$V_{5}\ (\mathbf{1},\ \mathbf{2},\ 3/2)$} & \textcolor{gray}{$V_{17}\ (\mathbf{3},\ \mathbf{2},\ 1/6)$} \\
\textcolor{gray}{$S_{15}\ (\mathbf{3},\ \mathbf{1},\ -1/3)$} & \textcolor{gray}{$F_{12}\ (\mathbf{3},\ \mathbf{2},\ -5/6)$} & \textcolor{gray}{$V_{17}\ (\mathbf{3},\ \mathbf{2},\ 1/6)$} \\
\textcolor{gray}{$S_{21}\ (\mathbf{3},\ \mathbf{2},\ 7/6)$} & \textcolor{gray}{$F_{12}\ (\mathbf{3},\ \mathbf{2},\ -5/6)$} & \textcolor{gray}{$V_{17}\ (\mathbf{3},\ \mathbf{2},\ 1/6)$} \\
\end{longtable}
\refstepcounter{table}
\begin{center}
\small\textbf{Table \thetable.} Short-range UV catalogue for $d_{\mathbb C}^\dagger e_{\mathbb C} L^\dagger Q u_{\mathbb C}^{2}$.
\label{tab:app_short_catalogue_3}
\end{center}
\endgroup

\begingroup
\small
\renewcommand{\arraystretch}{1.08}
\begin{longtable}{@{\hspace{1.2em}}>{\raggedright\arraybackslash}p{0.255\textwidth}@{\hspace{1.15em}}>{\raggedright\arraybackslash}p{0.255\textwidth}@{\hspace{1.15em}}>{\raggedright\arraybackslash}p{0.255\textwidth}@{}}
\hline
\multicolumn{3}{@{}l@{}}{{\normalsize\bfseries Operator type:} $d_{\mathbb C}^{\dagger 2} L^{\dagger 2} Q^{\dagger 2}$} \\
\hline
\textbf{Mediator 1} & \textbf{Mediator 2} & \textbf{Mediator 3} \\
\hline
\endfirsthead
\hline
\multicolumn{3}{@{}l@{}}{{\normalsize\bfseries Operator type:} $d_{\mathbb C}^{\dagger 2} L^{\dagger 2} Q^{\dagger 2}$} \\
\hline
\textbf{Mediator 1} & \textbf{Mediator 2} & \textbf{Mediator 3} \\
\hline
\endhead
\hline
\endfoot
$S_{15}\ (\mathbf{3},\ \mathbf{1},\ -1/3)$ & $S_{20}\ (\mathbf{3},\ \mathbf{2},\ 1/6)$ & $F^{(9)}_{5}\ (\mathbf{8},\ \mathbf{1},\ 0)$ \\
\cmidrule(lr){1-3}
$S_{15}\ (\mathbf{3},\ \mathbf{1},\ -1/3)$ & $S_{20}\ (\mathbf{3},\ \mathbf{2},\ 1/6)$ & $F_{13}\ (\mathbf{3},\ \mathbf{2},\ 1/6)$ \\
\cmidrule(lr){1-3}
$S_{15}\ (\mathbf{3},\ \mathbf{1},\ -1/3)$ & $S_{20}\ (\mathbf{3},\ \mathbf{2},\ 1/6)$ & $F_{1}\ (\mathbf{1},\ \mathbf{1},\ 0)$ \\
\cmidrule(lr){1-3}
$S_{15}\ (\mathbf{3},\ \mathbf{1},\ -1/3)$ & $S_{20}\ (\mathbf{3},\ \mathbf{2},\ 1/6)$ & $F_{24}\ (\mathbf{6},\ \mathbf{2},\ -1/6)$ \\
\cmidrule(lr){1-3}
$S_{15}\ (\mathbf{3},\ \mathbf{1},\ -1/3)$ & $S_{20}\ (\mathbf{3},\ \mathbf{2},\ 1/6)$ & $S_{39}\ (\mathbf{8},\ \mathbf{2},\ 1/2)$ \\
\cmidrule(lr){1-3}
$S_{15}\ (\mathbf{3},\ \mathbf{1},\ -1/3)$ & $S_{27}\ (\mathbf{6},\ \mathbf{1},\ -2/3)$ & $F_{12}\ (\mathbf{3},\ \mathbf{2},\ -5/6)$ \\
\cmidrule(lr){1-3}
$S_{15}\ (\mathbf{3},\ \mathbf{1},\ -1/3)$ & $S_{27}\ (\mathbf{6},\ \mathbf{1},\ -2/3)$ & $F_{24}\ (\mathbf{6},\ \mathbf{2},\ -1/6)$ \\
\cmidrule(lr){1-3}
$S_{15}\ (\mathbf{3},\ \mathbf{1},\ -1/3)$ & $S_{39}\ (\mathbf{8},\ \mathbf{2},\ 1/2)$ & $F^{(9)}_{5}\ (\mathbf{8},\ \mathbf{1},\ 0)$ \\
\end{longtable}
\refstepcounter{table}
\begin{center}
\small\textbf{Table \thetable.} Short-range UV catalogue for $d_{\mathbb C}^{\dagger 2} L^{\dagger 2} Q^{\dagger 2}$.
\label{tab:app_short_catalogue_4}
\end{center}
\endgroup

\begingroup
\small
\renewcommand{\arraystretch}{1.08}
\begin{longtable}{@{\hspace{1.2em}}>{\raggedright\arraybackslash}p{0.255\textwidth}@{\hspace{1.15em}}>{\raggedright\arraybackslash}p{0.255\textwidth}@{\hspace{1.15em}}>{\raggedright\arraybackslash}p{0.255\textwidth}@{}}
\hline
\multicolumn{3}{@{}l@{}}{{\normalsize\bfseries Operator type:} $d_{\mathbb C}^{\dagger 2} e_{\mathbb C} L^\dagger Q^\dagger u_{\mathbb C}$} \\
\hline
\textbf{Mediator 1} & \textbf{Mediator 2} & \textbf{Mediator 3} \\
\hline
\endfirsthead
\hline
\multicolumn{3}{@{}l@{}}{{\normalsize\bfseries Operator type:} $d_{\mathbb C}^{\dagger 2} e_{\mathbb C} L^\dagger Q^\dagger u_{\mathbb C}$} \\
\hline
\textbf{Mediator 1} & \textbf{Mediator 2} & \textbf{Mediator 3} \\
\hline
\endhead
\hline
\endfoot
$S_{15}\ (\mathbf{3},\ \mathbf{1},\ -1/3)$ & $S_{27}\ (\mathbf{6},\ \mathbf{1},\ -2/3)$ &  \\
\textcolor{gray}{$S_{15}\ (\mathbf{3},\ \mathbf{1},\ -1/3)$} & \textcolor{gray}{$S_{27}\ (\mathbf{6},\ \mathbf{1},\ -2/3)$} & \textcolor{gray}{$F^{(9)}_{3}\ (\mathbf{3},\ \mathbf{1},\ -4/3)$} \\
\textcolor{gray}{$S_{15}\ (\mathbf{3},\ \mathbf{1},\ -1/3)$} & \textcolor{gray}{$S_{27}\ (\mathbf{6},\ \mathbf{1},\ -2/3)$} & \textcolor{gray}{$F_{12}\ (\mathbf{3},\ \mathbf{2},\ -5/6)$} \\
\textcolor{gray}{$S_{15}\ (\mathbf{3},\ \mathbf{1},\ -1/3)$} & \textcolor{gray}{$S_{27}\ (\mathbf{6},\ \mathbf{1},\ -2/3)$} & \textcolor{gray}{$F_{23}\ (\mathbf{6},\ \mathbf{1},\ 1/3)$} \\
\textcolor{gray}{$S_{15}\ (\mathbf{3},\ \mathbf{1},\ -1/3)$} & \textcolor{gray}{$S_{27}\ (\mathbf{6},\ \mathbf{1},\ -2/3)$} & \textcolor{gray}{$F_{24}\ (\mathbf{6},\ \mathbf{2},\ -1/6)$} \\
\cmidrule(lr){1-3}
$F^{(9)}_{3}\ (\mathbf{3},\ \mathbf{1},\ -4/3)$ & $V_{16}\ (\mathbf{3},\ \mathbf{2},\ -5/6)$ & $V_{31}\ (\mathbf{8},\ \mathbf{1},\ 1)$ \\
\cmidrule(lr){1-3}
$F^{(9)}_{3}\ (\mathbf{3},\ \mathbf{1},\ -4/3)$ & $V_{2}\ (\mathbf{1},\ \mathbf{1},\ 1)$ & $V_{16}\ (\mathbf{3},\ \mathbf{2},\ -5/6)$ \\
\cmidrule(lr){1-3}
$F^{(9)}_{3}\ (\mathbf{3},\ \mathbf{1},\ -4/3)$ & $V_{2}\ (\mathbf{1},\ \mathbf{1},\ 1)$ & $V_{5}\ (\mathbf{1},\ \mathbf{2},\ 3/2)$ \\
\end{longtable}
\refstepcounter{table}
\begin{center}
\small\textbf{Table \thetable.} Short-range UV catalogue for $d_{\mathbb C}^{\dagger 2} e_{\mathbb C} L^\dagger Q^\dagger u_{\mathbb C}$.
\label{tab:app_short_catalogue_5}
\end{center}
\endgroup

\begingroup
\small
\renewcommand{\arraystretch}{1.08}
\begin{longtable}{@{\hspace{1.2em}}>{\raggedright\arraybackslash}p{0.255\textwidth}@{\hspace{1.15em}}>{\raggedright\arraybackslash}p{0.255\textwidth}@{\hspace{1.15em}}>{\raggedright\arraybackslash}p{0.255\textwidth}@{}}
\hline
\multicolumn{3}{@{}l@{}}{{\normalsize\bfseries Operator type:} $d_{\mathbb C}^{\dagger 2} e_{\mathbb C}^{2} u_{\mathbb C}^{2}$} \\
\hline
\textbf{Mediator 1} & \textbf{Mediator 2} & \textbf{Mediator 3} \\
\hline
\endfirsthead
\hline
\multicolumn{3}{@{}l@{}}{{\normalsize\bfseries Operator type:} $d_{\mathbb C}^{\dagger 2} e_{\mathbb C}^{2} u_{\mathbb C}^{2}$} \\
\hline
\textbf{Mediator 1} & \textbf{Mediator 2} & \textbf{Mediator 3} \\
\hline
\endhead
\hline
\endfoot
$F^{(9)}_{4}\ (\mathbf{3},\ \mathbf{1},\ 5/3)$ & $V_{13}\ (\mathbf{3},\ \mathbf{1},\ 2/3)$ & $V_{31}\ (\mathbf{8},\ \mathbf{1},\ 1)$ \\
\cmidrule(lr){1-3}
$F^{(9)}_{4}\ (\mathbf{3},\ \mathbf{1},\ 5/3)$ & $V_{2}\ (\mathbf{1},\ \mathbf{1},\ 1)$ & $V_{13}\ (\mathbf{3},\ \mathbf{1},\ 2/3)$ \\
\cmidrule(lr){1-3}
$F^{(9)}_{5}\ (\mathbf{8},\ \mathbf{1},\ 0)$ & $V_{13}\ (\mathbf{3},\ \mathbf{1},\ 2/3)$ & $V_{31}\ (\mathbf{8},\ \mathbf{1},\ 1)$ \\
\cmidrule(lr){1-3}
$F_{1}\ (\mathbf{1},\ \mathbf{1},\ 0)$ & $V_{2}\ (\mathbf{1},\ \mathbf{1},\ 1)$ & $V_{13}\ (\mathbf{3},\ \mathbf{1},\ 2/3)$ \\
\cmidrule(lr){1-3}
$S_{15}\ (\mathbf{3},\ \mathbf{1},\ -1/3)$ & $F^{(9)}_{3}\ (\mathbf{3},\ \mathbf{1},\ -4/3)$ & $V_{2}\ (\mathbf{1},\ \mathbf{1},\ 1)$ \\
\cmidrule(lr){1-3}
$S_{15}\ (\mathbf{3},\ \mathbf{1},\ -1/3)$ & $F^{(9)}_{3}\ (\mathbf{3},\ \mathbf{1},\ -4/3)$ & $V_{31}\ (\mathbf{8},\ \mathbf{1},\ 1)$ \\
\cmidrule(lr){1-3}
$S_{15}\ (\mathbf{3},\ \mathbf{1},\ -1/3)$ & $F^{(9)}_{5}\ (\mathbf{8},\ \mathbf{1},\ 0)$ & $V_{13}\ (\mathbf{3},\ \mathbf{1},\ 2/3)$ \\
\cmidrule(lr){1-3}
$S_{15}\ (\mathbf{3},\ \mathbf{1},\ -1/3)$ & $F^{(9)}_{5}\ (\mathbf{8},\ \mathbf{1},\ 0)$ & $V_{31}\ (\mathbf{8},\ \mathbf{1},\ 1)$ \\
\end{longtable}
\refstepcounter{table}
\begin{center}
\small\textbf{Table \thetable.} Short-range UV catalogue for $d_{\mathbb C}^{\dagger 2} e_{\mathbb C}^{2} u_{\mathbb C}^{2}$.
\label{tab:app_short_catalogue_6}
\end{center}
\endgroup

\bibliographystyle{JHEP}
\bibliography{references}

\end{document}